\documentclass[aps,prc,showpacs,preprint]{revtex4-1}

\usepackage{graphics,epsfig}
\usepackage{amsmath}
\usepackage{rotating}

\newcommand {\bx}[1]  {\mbox{\boldmath $#1$}}
\newcommand{\beq}{\begin{equation}}
\newcommand{\eeq}[1]{\label{#1}\end{equation}}
\newcommand{\beqa}{\begin{eqnarray}}
\newcommand{\eeqa}[1]{\label{#1}\end{eqnarray}}
\newcommand{\eeqan}{\end{eqnarray}}

\begin{document}

\title{Effective model for in-medium $\bar{K}N$ interactions including the $L=1$ partial wave}

\author{Ale\v{s} Ciepl\'y$^1$}
\email{cieply@ujf.cas.cz}
\affiliation{Nuclear Physics Institute, Academy of Sciences of the Czech Republic,
\v{R}e\v{z}, Czech Republic}

\author{Vojt\v{e}ch Krej\v{c}i\v{r}\'{i}k$^2$}
\email{vojtech.krejcirik@riken.jp}
\affiliation{Theoretical Research Division, RIKEN Nishina Center, Wako, Saitama, Japan}

\begin{abstract}

Coupled channels model of meson-baryon interactions based on the effective chiral Lagrangian is extended 
to account explicitly for the $\Sigma(1385)$ resonance that dominates the $P$-wave $\bar{K}N$ and $\pi\Sigma$ 
interactions at energies below the $\bar{K}N$ threshold. The presented model aims at a uniform treatment 
of the $\Lambda(1405)$ and $\Sigma(1385)$ dynamics in the nuclear medium. We demonstrate the applicability 
of the model by confronting its predictions with the vacuum scattering data, then we follow with discussing 
the impact of nuclear matter on the $\pi\Sigma$ mass spectrum and on the energy dependence 
of the $K^{-}p$ branching ratios.

\end{abstract}


\maketitle

\section{Introduction}
\label{sec:intro}

The low energy interactions of anti-kaons with nucleons represent a vital area where QCD, the quantum field theory of strong 
interactions, is tested in its nonperturbative region. 
For energies close to the threshold, the $\bar{K}N$ interactions are dominated by the $\Lambda(1405)$ and $\Sigma(1385)$ 
resonances in the $S$ and $P$ waves, respectively. There, one often resorts to effective field theories \cite{1979Wei}, 
\cite{1985GL} combined with coupled channels resummation techniques \cite{1995KSWnpa}, \cite{1998OR} to keep the QCD 
symmetries and overcome problems with convergence of the perturbative series in the presence of resonances. 
The sizable attraction existing between the anti-kaon 
and the nucleon is also relevant for a likely existence of the three-body $K^{-}pp$ ``molecular state'' 
(see e.g. \cite{2013MAY} and references there), for a formation of quasi-bound (anti)kaon-nuclear states 
\cite{2006MFG}, \cite{2012GM}, or for a description of a hot and dense matter realized in heavy-ion collisions 
or in compact stars \cite{2008SSS}. 

The nature and properties of strangeness $S=-1$ baryon resonances are also very important issues in hadronic physics. 
The $\Sigma(1385)$ with spin-parity $J^{P} = 3/2^{+}$ is considered to belong to the baryon decuplet as a first excited 
state of $\Sigma(1193)$ - this is in agreement with insight coming from the analysis of QCD in the limit of infinite 
number of colors \cite{1974tH}, \cite{1979Wit}. In the $N_c \rightarrow \infty$ world, consistency relations 
\cite{1984GSprl}, \cite{1993DM-1} require the decuplet of resonances involving $\Sigma(1385)$
to be a part of a degenerate ground state multiplet together with the octet baryons. On the other hand 
the traditional quark picture has difficulty to accommodate the $\Lambda(1405)$ ($J^{P} = 1/2^{+}$) 
which is more likely a molecular $\bar{K}N$ state rather than an excited three quarks state \cite{2012HJ}. 
Currently, the most popular view is that $\Lambda(1405)$ is generated dynamically as a $\bar{K}N$ quasi-bound state 
submerged in $\pi\Sigma$ continuum that provides another much broader resonance \cite{2008HW}. This picture emerges from coupled 
channels analysis with the inter-channel dynamics 
derived from an effective chiral Lagrangian. As a result one gets 
two resonant states close to the $\bar{K}N$ threshold that are assigned to the $\Lambda(1405)$ \cite{2001OM}, \cite{2003JOORM}. 
The state at a higher energy, typically around 1420-1430 MeV, couples more strongly to the $\bar{K}N$ channel and has 
a moderate width of 40-80 MeV, while the position of the other state is more model dependent and acquires 
a much broader width. 

The formation of $\Lambda(1405)$ has been a subject of several recent experiments aiming at establishing its structure. 
Provided the resonance is composed of two states, the observed mass spectra should depend on the reaction mechanism, 
particularly on involvement of either $\pi\Sigma$ or $\bar{K}N$ channels to which the two states couple with 
a different strength. Although this hypothesis explains the asymmetric shape of the $\Lambda(1405)$ 
and generally is believed to be correct, the experimental results are still puzzling. The $\Lambda(1405)$ 
signal in the $pp \longrightarrow Y^{\ast} + P + K^{+}$ reaction measured by the HADES collaboration 
at GSI \cite{2013HADES} is compatible with the two-pole structure of the $\Lambda(1405)$ as well as with 
the earlier measurement by the ANKE collaboration at COSY \cite{2008ANKE} or with the analysis 
of much older bubble chamber data from the $\pi^{-}p$ experiments \cite{1984Hem}. On the other hand, 
the $\pi\Sigma$ mass spectra measured in a photoproduction reaction on a proton by the CLAS collaboration 
at JLab \cite{2013CLAS} have the $I=0$ peak at too low energy, 
close to the $\pi\Sigma$ threshold. Nevertheless, the recent theoretical analyses 
\cite{2013RO}, \cite{2015MM} seem to succeed at accomodating the data within the standard 
chirally motivated models. In addition two $I=1$ peaks were observed in the CLAS experiment 
at energies around 1400 MeV contributing to the enigmatic nature of the physics 
in the vicinity of the $\Lambda(1405)$ resonance.

Another source of information on the $\Lambda(1405)$ and $\Sigma(1385)$ resonances should come from the studies of 
kaon-nuclear clusters by the AMADEUS experiment performed in Frascati \cite{2013AMAD}. As the preliminary data show, it is tricky 
to disentangle the contributions of the two excited hyperonic states when nuclear matter effects add up to other 
experimental uncertainties. Typically, the $\Lambda(1405)$ mass spectra extracted from the $\pi^{\pm}\Sigma^{\mp}$ 
distributions are contaminated by the $\Sigma(1385)$ continuum. There, a theoretical model involving in-medium 
dynamics of both resonances treated in a uniform way might be of a vital help. In the present paper we aim at 
developing such model by using chirally motivated effective meson-baryon potentials that are contrained 
by the available experimental data on low-energy $\bar{K}N$ interactions.

In the free space $S$-wave $\bar{K}N$ and $\pi\Sigma$ systems are reliably well described by chirally motivated 
coupled channels models that involve interactions of the $J^P = 0^{-}$ meson octet with the $J^P = 1/2^{+}$ baryon octet
\cite{1995KSWnpa}, \cite{1998OR}, \cite{2003JOORM}, \cite{2005BNW}, \cite{2012CS}, \cite{2012IHW}, \cite{2013GO}, \cite{2015MM}. 
Several authors have also extended the vacuum model to incorporate the influence of nuclear matter on the $\bar{K}N$ system
\cite{1996WKW}, \cite{1998Lut}, \cite{2000RO}, \cite{2001CFGM}, \cite{2006TRO}, \cite{2011CFGGM}. It has been demonstrated 
that the dynamics of the $\Lambda(1405)$ is responsible for switching the sign of the $\bar{K}N$ scattering length 
from a negative value to a positive one at a relatively small density, about 1/10 of the typical nuclear 
density $\rho_0 = 0.17$ fm$^{-3}$ \cite{1994FGB}. The $K^{-}$-nuclear interaction becomes even more attractive 
at subthreshold energies probed by kaonic atoms, though even larger attraction (most likely due to $\bar{K}NN$ interactions) 
is needed to comply with phenomenological analysis of the experimental data \cite{2011CFGGM}, \cite{2013FG}. 
A downside of these models, however, is that they are restricted only to the lowest partial wave $L=0$ 
leaving certain aspects of physics completely inaccessible---angular distribution of scattering 
cross section being the most obvious one. There were attempts to extend the approach to $P$-wave interactions 
in \cite{2002JOR} and \cite{2012Kre}, and the first model was also used to study the properties of antikaons 
in nuclear matter \cite{2006TRO}. Since the $\Sigma(1385)$ is not generated dynamically \cite{2012Kre}, 
in the present paper we build on an already established $S$-wave model of Ref.~\cite{2012CS} by adding 
the $\Sigma(1385)$ resonance explicitly. As this resonance completely dominates the $P$-wave 
interactions around the $\bar{K}N$ threshold an introduction of additional $P$-wave contributions does not seem 
necessary. With the model established to reproduce the available experimental data we then proceed to study 
the impact of nuclear medium on the shape of the $\pi\Sigma$ mass spectra. In addition 
we also look at density and energy dependence of several $K^{-}N$ branching ratios.

The paper is organized as follows. In section \ref{sec:formalism}, the general formalism of the 
spin 0 -- spin 1/2 scattering is reviewed. The model itself is formulated in section \ref{sec:potentials}   
where its vacuum version as well as an extension to nuclear medium are presented. Our results and 
conclusions are discussed in sections \ref{sec:results} and \ref{sec:conc}, respectively.

\section{Potential scattering formulation}
\label{sec:formalism}

The main purpose of this section is to set the notation and introduce the terms and concepts 
related to the scattering of two particles, one of which has spin one-half and the other one 
is spinless (as is the case of meson-baryon scattering). The analysis presented here is restricted 
to interactions that are both time reversal and parity invariant. This loss of generality is fully 
justified since our ultimate goal is to understand processes governed by the strong interaction, 
i.e. the interaction that is invariant with respect to these symmetries. Additionally, throughout 
the paper only the first two partial waves ($L=0$ and $L=1$) are taken into account.

\subsection{General formalism of spin-zero spin-half scattering}

The most general form of the scattering amplitude (recall that the same decomposition 
can be written down for the potential and, in general, for any quantity that is
time reversal and parity invariant) is a $2 \times 2$ matrix in the spin space \cite{1964GW}:
\begin{equation}
F(\bx{p} \rightarrow \bx{p'}) = \tilde{F}(\bx{p} \rightarrow \bx{p'})
+ {\rm i} \, \bx{\sigma} \cdot \bx{\hat{p}} \times \bx{\hat{p}'} \,\, \tilde{G}(\bx{p} \rightarrow \bx{p'}) \;.
\label{eq:amplgeneral}
\end{equation}
Considering only the $S$ and $P$ partial waves,
one obtains three independent components with well defined parity and total
angular momentum:
\begin{itemize}
\item $L=0$, $J=1/2$ : $0+$ partial wave,
\item $L=1$, $J=1/2$ : $1-$ partial wave,
\item $L=1$, $J=3/2$ : $1+$ partial wave.
\end{itemize}
The spin-flip ($\tilde{G}$) and and spin-non-flip ($\tilde{F}$) amplitudes of 
Eq.~(\ref{eq:amplgeneral}) can be written in terms of the partial wave components as
\begin{equation}
\begin{aligned}
\tilde{F} &=  F^{0+} + \left( 2F^{1+} + F^{1-} \right) \bx{\hat{p}} \cdot \bx{\hat{p}'} \;,\\
\tilde{G} &=  F^{1+} - F^{1-}  \;,
\end{aligned}
\end{equation}
where the hat indicates a unit vector in the respective direction.
The general formula for the differential cross section with
unpolarized beam and target is given as
\begin{equation}
\frac{{\rm d}\sigma}{{\rm d}\Omega} (\bx{p} \rightarrow \bx{p'})
= |\tilde{F}(\bx{p} \rightarrow \bx{p'})|^2 + \sin^2(\theta) \,\, |\tilde{G}(\bx{p} \rightarrow \bx{p'})|^2 \;.
\end{equation}
If only $L=0$ and $L=1$ are included --- the case in which we are interested --- the
differential and total cross sections read:
\begin{align}
\frac{{\rm d}\sigma}{{\rm d}\Omega} &= \left|F^{0+}\right|^2 + \left|2F^{1+}+F^{1-}\right|^2 \cos^2\theta 
                                     + \left|F^{1+}-F^{1-}\right|^2 \sin^2\theta + \nonumber \\
                                    &+ \left( F^{0+}(2F^{1+} + F^{1-})^* +(F^{0+})^*(2F^{1+} + F^{1-}) \right) \cos\theta 
                                        , \label{sigmaforL01} \\
\sigma^{\rm tot} &= 4\pi \left( \left|F^{0+}\right|^2 + \frac{1}{3} \left|2F^{1+}+F^{1-}\right|^2 + \frac{2}{3} \left|F^{1+}-F^{1-}\right|^2\right)  .
\end{align}

The two-body potential obeying the desired symmetries (parity and time reversal) can be decomposed 
in exactly the same way --- into three independent partial-wave components ($V^{0+}$, $V^{1+}$, and $V^{1-}$),
\begin{equation}
V = V^{0+} + \left( 2 V^{1+} + V^{1-} \right) \bx{\hat{p}} \cdot \bx{\hat{p}'} + 
\left( V^{1+} - V^{1-} \right) {\rm i}\, \bx{\sigma} \cdot \bx{\hat{p}} \times \bx{\hat{p}'}  \;.
\label{eq:potentialdecomposition}
\end{equation}
In this scheme, the Lippmann-Schwinger equation for scattering amplitude splits into
three independent equations --- one for each partial wave ($0+$, $1-$, and $1+$) --- with
angular dependence fully determined by Eqs.~(\ref{eq:amplgeneral}), and (\ref{eq:potentialdecomposition}):
\begin{equation}
F^{0+/1-/1+} = V^{0+/1-/1+} + V^{0+/1-/1+} G^{0/1} F^{0+/1-/1+} \;.
\end{equation}

\subsection{The Lippmann-Schwinger equation with separable potentials}

Considering only the $S$ and $P$ waves, there are three partial-wave effective potentials 
taken in the separable form, i.e.
\begin{equation}
\begin{aligned}
V_{ij}^{0+}(p, p') & = g_{i}^0(p) \,\, v_{ij}^{0+}  \,\, g_{j}^0(p') \;,\\
V_{ij}^{1-}(p, p') & = g_{i}^1(p) \,\, v_{ij}^{1-}  \,\, g_{j}^1(p') \;,\\
V_{ij}^{1+}(p, p') & = g_{i}^1(p) \,\, v_{ij}^{1+}  \,\, g_{j}^1(p') \;.
\end{aligned}
\end{equation}
The indexes $i$ and $j$ tag specific meson-baryon channels in the initial and final states, respectively,
and $g_{j}^L (p)$ stands for form factors in a given partial wave (and, in principle, 
for each scattering channel involved). They carry all the information about the incoming and outgoing
momenta, more precisely about the absolute values of momenta, since the angular
dependence has been factorized in the previous section. The form factors are taken 
in the Yamaguchi form \cite{1954Yam}, \cite{1954YamYam} as
\begin{align}
g_{j}^0 (p) & = 1\left/\middle(1+\frac{p^2}{\alpha_{j}^2}\right)    \;, \label{eq:ffYam0}\\
g_{j}^1 (p) & = p\left/\middle(1+\frac{p^2}{\alpha_{j}^2}\right)^{3/2}  . \label{eq:ffYam1}
\end{align}
Here $\alpha_{j}$ represents the inverse range parameter that is (in general) channel dependent.
A concrete choice of form factors is arbitrary to a point. However, the Yamaguchi 
form for the $S$-wave (\ref{eq:ffYam0}) is widely used 
within the community and the form for $P$-wave (\ref{eq:ffYam1}) is its very natural extension.
In particular, the form of $g^1 (p)$ is constrained by two requirements: first, the amplitude
must satisfy a general low-momentum condition $f=p^{2L+1}\cot(\delta)$ for $p\rightarrow 0$ (this determines
the linear dependence on $p$); second, we want the convergence of the Green's functions 
in the Lippmann-Schwinger equation to be of the same order for both partial waves, 
i.e.~$g^0/g^1 \rightarrow {\rm const.}$ when $p\rightarrow \infty$ (this determines the power 
of the denominator in the $g^1$ form). Let us note that the systematic error 
introduced at this point is compensated by the fact that the inverse ranges 
$\alpha_j$ are free parameters of the model that will eventually be fitted to data.

The separable form of the potentials allows us to use the same ansatz for the
scattering amplitudes as well,
\begin{widetext}
\begin{equation}
\begin{aligned}
F_{ij}^{0+}(p, p') & = g_{i}^0(p) \,\, f_{ij}^{0+}  \,\, g_{j}^0(p') \;,\\
F_{ij}^{1-}(p, p') & = g_{i}^1(p) \,\, f_{ij}^{1-}  \,\, g_{j}^1(p') \;,\\
F_{ij}^{1+}(p, p') & = g_{i}^1(p) \,\, f_{ij}^{1+}  \,\, g_{j}^1(p') \;.
\end{aligned}
\end{equation}
Then, the Lippmann-Schwinger equation is reduced into three algebraic equations,
\begin{equation}
\begin{aligned}
f_{ij}^{0+} & = \left[ \left( \mathbf{1} - v^{0+} \cdot G^0 \right)^{-1}  \cdot v^{0+} \right]_{ij} \;,\\
f_{ij}^{1-} & = \left[ \left( \mathbf{1} - v^{1-} \cdot G^1 \right)^{-1}  \cdot v^{1-} \right]_{ij} \;,\\
f_{ij}^{1+} & = \left[ \left( \mathbf{1} - v^{1+} \cdot G^1 \right)^{-1}  \cdot v^{1+} \right]_{ij} \;,
\end{aligned}
\label{eq:LSinPWD}
\end{equation}
where the Green functions are energy dependent diagonal matrices represented by the integrals
\begin{align}
G^0_{n}(\sqrt{s}) & = -\frac{1}{2\pi^2} \int {\rm d}^3q \frac{ \left[g^0(q) \right]^2}{p_{n}^2 - q^2 + {\rm i} \epsilon}  
                    = \frac{\alpha}{2} \,\frac{\alpha^2}{\left(\alpha - {\rm i} p_{n}\right)^2} \;,\label{eq:GreenF0}\\
G^1_{n}(\sqrt{s}) & = -\frac{1}{2\pi^2} \int {\rm d}^3q \frac{ \left[g^1(q) \right]^2}{p_{n}^2 - q^2 + {\rm i} \epsilon}  
                    = \frac{\alpha^{3}}{2}\, \frac{\alpha^2\left(\alpha-3 {\rm i} p_{n}\right)}{\left(\alpha - {\rm i} p_{n}\right)^3} \;. 
\label{eq:GreenF1}
\end{align}
\end{widetext}
Here $p_n$ denotes the on-shell meson-baryon CMS momenta in the $n$-th channel and the integration 
can be performed analytically in the free space.

\section{Chirally and colorfully motivated potentials}
\label{sec:potentials}

In this section, the chiral and large $N_c$ based model for the meson-baryon interactions in
the strange sector that includes both $S$- and $P$-waves is formulated. Our aim is to extend 
the existing model of Ref.~\cite{2012CS}, which includes only the $S$-wave,
to accommodate the $P$-wave physics as well.

The dynamics of the anti-kaon nucleon system is determined by the coupled channels 
potentials within the formalism presented in the previous section.
We employ potentials that are motivated by the chiral $SU(3)$ dynamics and
comply with the large $N_c$ consistency relations.
For the $L=0$ part of our model we take the leading order Tomozawa-Weinberg (TW) interaction 
that is represented by potentials used successfully and discussed in detail in \cite{2012CS}. 
Here, we rather focus on the $P$-wave physics, the main novelty of the current work.

\begin{figure}
\includegraphics[scale=0.4]{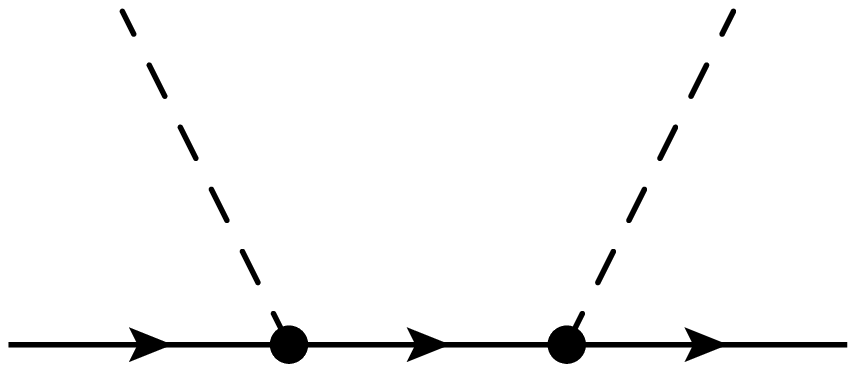}\hspace{0.5cm}
\includegraphics[scale=0.4]{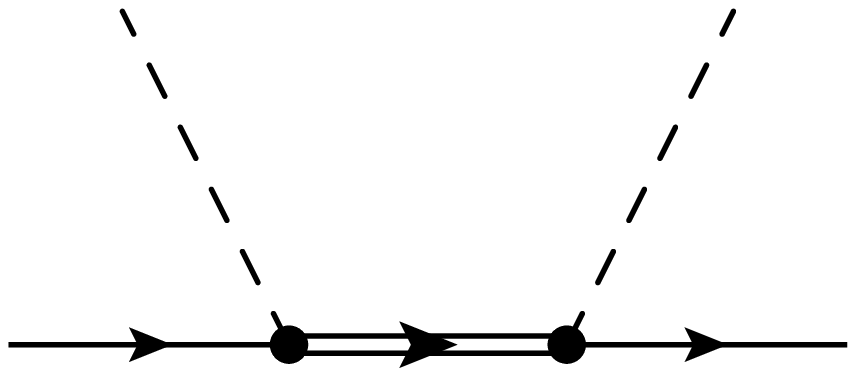}\\\vspace{0.5cm}
\includegraphics[scale=0.4]{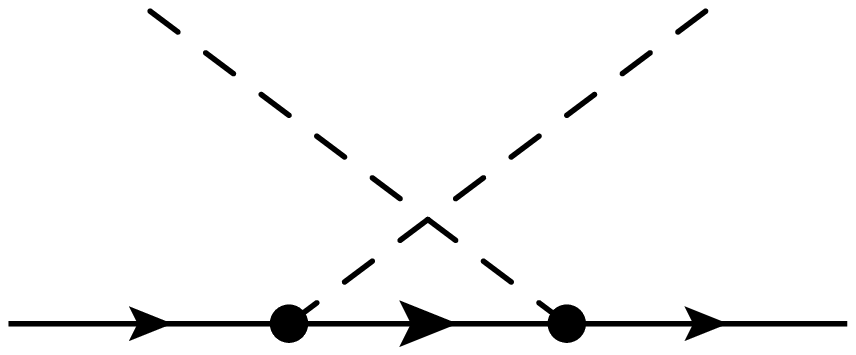}\hspace{0.5cm}
\includegraphics[scale=0.4]{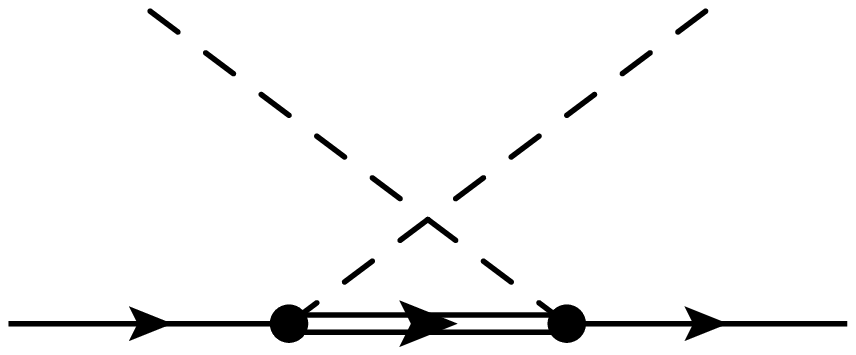}
\caption{Feynman diagrams representing leading order interactions in the $P$-wave sector.}
\label{fig:diagrams}
\end{figure}

It is well known that the $S$-wave $\Lambda(1405)$ resonance is generated dynamically already 
by the leading order TW interaction. On the other hand, the situation is principally different in the $L=1$ sector, 
where the most significant contributions come from the direct and crossed Born diagrams. It was discussed 
in the preceding work \cite{2012Kre} that considering only octet baryons in the intermediate
state (left diagrams in Fig.~\ref{fig:diagrams}), one is not able to reproduce the resonant behavior 
which could be identified as the $\Sigma(1385)$. Thus, in order to formulate a phenomenologically 
relevant model, one is forced to include the decuplet baryons with the $\Sigma(1385)$ being one of them, 
as building blocks as well (right diagrams in Fig.~\ref{fig:diagrams}). We recall that this situation 
(treating both octet and decuplet baryons in the same way) is analogous to the one which occurs 
in the large $N_c$ limit of QCD. There, the consistency relations \cite{1984GSprl}, \cite{1993DM-1}, 
\cite{1984GSprd}, \cite{1993DM-2}, \cite{1994DJM}, \cite{1995DJM} require the octet and decuplet baryons to 
be degenerate and form a ground state multiplet of the contracted $SU(2N_f)$ symmetry. It seems natural 
to take inspiration from this idea when designing the $P$-wave part of our model.

Having decuplet baryons as fundamental degrees of freedom, 
one can construct four different diagrams contributing in the leading
order: a direct one with the octet baryon in the intermediate state, a direct one
with the decuplet baryon in the intermediate state, a crossed one with 
the octet baryon in the intermediate state, and a crossed one
with the decuplet baryon in the intermediate state.
These four possibilities are depicted in Fig.~\ref{fig:diagrams}.

In general, the meson-baryon-baryon vertex is of the form
\begin{equation}
\frac{\partial^i \phi^a }{f_{\pi}} \,\, \left\langle B'| \, \sigma^i \, \lambda^a \, |B \right\rangle  \,\, \sim \,\, X^{ia} \;,
\label{eq:MBBvertex}
\end{equation}
where the $X^{ia}$ stands for a generalized spin-flavor operator acting on the 
baryon ground state multiplet \cite{1994DJM}, \cite{1995DJM}.
It represents a straightforward generalization of a $\sigma^i\tau^i$ Pauli matrix product 
standing at a standard pion-nucleon vertex.
This operator is strongly constrained by the emergent spin-flavor symmetry at large $N_c$.
Specifically, the relative strengths of the couplings are determined by appropriate Clebsch-Gordan 
coefficients of $SU(2)$-spin and $SU(3)$-flavor group,
\begin{widetext}
\begin{align}
\Gamma \sim & \, \left( {\rm baryon IN}, {\rm meson IN} | {\rm baryon OUT} \right)_{SU(2)-{\rm spin}} \nonumber\\
&\left( {\rm baryon IN}, {\rm meson IN} | {\rm baryon OUT} \right)_{SU(3)-{\rm flavor}} \; ,
\end{align}
\end{widetext}
and the absolute strengths are fixed to obey the the large $N_c$ consistency relations. 
The relevant $SU(3)$-flavor Clebsch-Gordan coefficients as well as absolute strengths
of the couplings are discussed in detail in the Appendix, and the coupling matrices for the $SU(2)$-spin
transitions can be found in \cite{2012CK}.

Let us note that even though we use the large $N_c$ limit to motivate the model,
the Clebsch-Gordan coefficients for $SU(3)$-flavor group considered in this paper
correspond to $N_c=3$ \cite{1963dS}, \cite{2004CL}. The reason for this is a practical one, particularly 
because the Clebsch-Gordan coefficients including states with strangeness are suppressed 
as $N_c\rightarrow\infty$ thus leading to a trivial interaction.

\subsection{Free space potential}

The construction of the $S$-wave effective potential by matching it to the chiral meson-baryon amplitude 
is well documented in the literature \cite{1995KSWnpa}, \cite{1998OR}, \cite{2012CS}. For the $0^+$ partial wave, 
the leading order contribution comes from the Tomozawa-Weinberg term which we take in a form
\begin{align}
V^{TW}_{ij}(p, p') & = -\frac{1}{4\pi f_{\pi}^2} \sqrt{\frac{M_i M_j}{s}} \left[ -\frac{1}{4} \, \mathcal{C}^{TW} \left( \sqrt{s} - M_i - M_j \right) \right]  \;.
\label{eq:TW}
\end{align}
In the present work it is the only potential contribution we consider in the $S$-wave and since it bears 
no angular dependence, $V^{TW}_{ij}(p, p')$ is also equal to $v_{ij}^{0+}$. The strength of the interchannel 
couplings is determined by the $\mathcal{C}^{TW}$ matrix which is given by the standard SU(3) Clebsch-Gordan coefficients.
In principle, the potential (\ref{eq:TW}) should be complemented by contributions that represent higher orders in the chiral 
expansion, and by the direct and crossed Born diagrams. However, since we concentrate on the impact of the $P$-wave and on 
in-medium applications we prefer to keep the meson-baryon potential as simple as possible and leave a further refinement 
of our model for a future.

In the $P$-wave sector, it is straightforward to determine which of the four leading-order diagrams shown 
in Fig.~\ref{fig:diagrams} plays a key role in the energy region of interest. Due to a suppression 
in the energy denominator caused by the necessity to excite the incoming baryon, the dominant contribution 
comes from the direct Born term with the decuplet in the intermediate state. For the strangeness-isospin 
combination discussed in this study, the intermediate baryon is the $\Sigma(1385)$. 
Thus, our $P$-wave potential can be written as
\beq
V_{ij}(p, p') = -\frac{1}{4\pi f_{\pi}^2} \sqrt{\frac{M_i M_j}{s}} \,\,\,\, \mathcal{C}^{d10}_{\Sigma^{*0}} 
\left(- \frac{\sqrt{2}}{\sqrt{s} - \tilde{M}_{\Sigma^{*0}}} \,\, \bx{p} \cdot \bx{p'} -\frac{\frac{1}{\sqrt{2}}}{\sqrt{s} -  \tilde{M}_{\Sigma^{*0}} } \,\, {\rm i}\, \bx{\sigma} \cdot \bx{p} \times \bx{p'} \right)  \; ,
\eeq{eq:Vdir10}
where $\tilde{M}_{\Sigma^{*0}}$ represents the bare $\Sigma(1385)$ mass.
From here, one immediately obtains the relevant partial wave component that enters the Lippmann-Schwinger 
equation (\ref{eq:LSinPWD}), 
\beq
v^{1+}_{ij} = -\frac{1}{4\pi f_{\pi}^2} \sqrt{\frac{M_i M_j}{s}}  
\left( -\frac{1}{\sqrt{2}} \,\mathcal{C}^{d10}_{\Sigma^{*0}}\, \frac{1}{\sqrt{s} - \tilde{M}_{\Sigma^{*0}}} \right) \;.
\eeq{eq:V1+}
The coupling matrix $\mathcal{C}^{d10}_{\Sigma^{*0}}$ is determined 
by the $SU(3)$-flavor Clebsch-Gordan coefficients and large $N_c$ consistency
relations following our motivation at the beginning of this section. The exact specification 
of the $\mathcal{C}^{d10}_{\Sigma^{*0}}$ matrix is relayed to the Appendix.
We also note that restricting our $P$-wave only to the excitation of $\Sigma^*$, 
which has a total angular momentum $3/2$, implies that the partial wave potential component $v^{1-}$ 
is identically zero.

\subsection{In-medium extension}
\label{sec:medium}

As mentioned in Section \ref{sec:intro} and already tested in several applications restricted 
to the $S$-wave sector, the separable potential model is well suited to study the impacts of nuclear matter on meson-baryon 
interactions. The analysis of nuclear-medium effects on the full model (including the $P$-wave) 
is presented in this section.

In principle, the nuclear medium affects the meson-baryon system in two ways. 
Firstly, for the channels involving nucleons the Pauli principle restricts the available phase-space 
which effectively shifts the respective intermediate state threshold to higher energies. 
Secondly, the interacting hadrons acquire self-energies due to their interaction with nuclear 
matter which shifts the position of the pole in the intermediate state propagator. 
For the $\bar{K}N$ system both effects compensate each other to a large extent \cite{2011CFGGM}. 
The Pauli blocking moves the $\Lambda(1405)$ structure above the $\bar{K}N$ threshold
where the resonance is partly dissolved \cite{1996WKW}, but the inclusion of nucleon and kaon 
self-energies \cite{1998Lut} brings it back below the threshold, though the $\bar{K}N$ related 
pole apparently moves to a more distant Riemann sheet. 

When implementing the nuclear medium effects we follow the procedure adopted in Refs.~\cite{2012CS} 
and \cite{2011CFGGM}. The intermediate state meson-baryon Green function of Eqs.~(\ref{eq:GreenF0}) 
and (\ref{eq:GreenF1}) is then modified as:
\beq
G_{n}^{L}(\sqrt{s},\rho) = -4\pi \: \int_{\Omega_{n}(\rho)} \frac{{\rm d}^{3}q}{(2\pi)^{3}}
\frac{[g_{n}^{L}(q^{2})]^{2}}
     {p_{n}^{2}-q^{2} -\Pi_{n}(\sqrt{s},\rho) +{\rm i}\epsilon}  \;,
\eeq{eq:Grho}
with $\Pi_{n}(\sqrt{s},\rho)$ standing for the sum of the meson and baryon self-energies \cite{2001CFGM}. 
The integration is carried over the $\Omega_{n}(\rho)$ domain of momenta allowed by the Pauli 
principle (intermediate state nucleon momenta above the Fermi limit) \cite{1996WKW}.
In such situation, the integral in Eq.~(\ref{eq:Grho}) is to be evaluated numerically.
We also note that not only the Green function (\ref{eq:Grho}) but also the amplitudes $f_{ij}$ 
that one gets by solving Eqs.~(\ref{eq:LSinPWD}) become density dependent quantities. 

To simplify the matters we consider the nucleons at rest when evaluating the domain $\Omega_{n}(\rho)$, 
neglecting completely the influence of their Fermi motion. With the exception of kaons that are treated 
selfconsistently we also assume a simple energy independent form for the hadron self-energies 
\beq
\Pi^{h}_{n}(\rho) = 2\mu_{n}\, V^{h}_{0}\, \rho / \rho_{0} \; ,
\eeq{eq:SE}
where $h$ tags the specific particle (either meson or baryon) and $\mu_{n}$ is the relativistic 
meson-baryon reduced energy. For the baryons we take $V_{0}^{\Lambda} = (-30-\rm{i}10)$ MeV, 
$V_{0}^{\Sigma} = (30-\rm{i}10)$ MeV and $V_{0}^{N} = (-60-\rm{i}10)$ MeV, 
for the optical potential depths, the values that are consistent with mean-field potentials 
used in nuclear structure calculations and were already tested in previous $\bar{K}$-nuclear 
and $\eta$-nuclelar calculations \cite{2011CFGGM}, \cite{2014CFGM}. 
The imaginary parts of the baryon optical potentials reflect a possibility 
of the inelastic hadron-nuclear processes as well as of the $\Sigma N \rightarrow \Lambda N$ 
conversion. The uniformly adopted value of $\Im V_0 = -10$ MeV represents a fair 
estimate for all three hadrons with any anticipated corrections falling below  
the overall level of theoretical ambiguities present in our model.
We also note in passing that the self-energies that enter Eq.~(\ref{eq:Grho}) differ 
by a kinematic factor $\mu_{n} /E_{h}$ (with $E_h$ denoting the hadron energy in the meson baryon CMS) 
from the self-energies used in the equation of motion written in the laboratory frame where the nuclear 
matter is at rest. 

The evaluation of the pion optical potential depth $V^{\pi}_{0}$ at energies in a vinicity of the $\Delta(1232)$ 
resonance represents a delicate issue. There, the $\pi N$ interaction changes from attractive below 
the resonance to repulsive at energies above the resonance. For the $\pi \Sigma$ CMS energies below 
the $\bar{K}N$ threshold the pion momenta are about $130-180$ MeV in the LAB system with nuclear matter 
at rest. This relates to the $\pi N$ CMS energies below the $\Delta(1232)$ resonance, where 
the pion-nuclear optical potential is attractive and few tens of MeV deep at central nuclear density \cite{1996JS}.
However, when the Fermi motion and other nuclear medium effects are considered, it is clear 
that our interactions cover a broad region of $\pi N$ energies below and above the $\Delta(1232)$ 
resonance. Thus, it is feasible to approximate the pion optical potential with a purely imaginary value 
of $V^{\pi}_{0}$. The later quantity can be estimated by using a relation
\beq
2 \mu_{\pi N}\, \Im V^{\pi}_{0}(\sqrt{s}) = -4 \pi\, \Im F_{\pi N}(\sqrt{s}, 0^{\circ})\, \rho_{0} 
= - p_{\pi N}\, \sigma_{\pi N}(\sqrt{s})\, \rho_0
\eeq{eq:vpi}
where $F_{\pi N}(\sqrt{s}, 0^{\circ})$ represents a forward free space scattering amplitude 
and $\sigma_{\pi N}$ stands for the total pion-nucleon cross section. At the energies related to our work 
the SAID database \cite{SAID} provide the $F_{\pi N}$ amplitudes (or cross sections) 
that result into $\Im V^{\pi}_{0}$ values from $-10$ to $-50$ MeV. For the purpose of our analysis, 
we have decided to go with $V^{\pi}_{0} = -{\rm i}30$ MeV, a value consistent with 
the phenomenological potentials reported in Refs. \cite{1996JS}, \cite{1983F}.

Finally, the (anti)kaon optical potential is standardly constructed 
as a coherent sum of one-nucleon contributions which leads to the selfenergy 
\beq
\Pi^{\bar{K}} = -4\pi\: F_{\bar{K}N} \: \rho       \; .
\eeq{eq:SEkaon}
Since this self-energy is expressed through the $\bar{K}N$ amplitudes that 
are obtained as a solution of the Lipmann-Schwinger equations (\ref{eq:LSinPWD}) 
the coupled channels system of equations has to be solved iteratively
to achieve selfconsistency, see Refs.~\cite{2001CFGM}, and \cite{2011CFGGM} for details. 
Standardly, only 5-7 iterations are sufficient in the procedure.

\section{Free space and in-medium results}
\label{sec:results}

The effective potential constructed in previous section contains several free parameters 
that can be fixed in fits to low energy $K^{-}p$ data. Here we adopt an already established
TW1 parametrization of \cite{2012CS} for the $0+$ potential with only one inverse range 
used in all channels, $\alpha_{j}=\alpha=701$ MeV, and with the meson decay constant 
$f_{\pi}=113$ MeV. With this simple setting one gets quite satisfactory description 
of all available low energy $K^{-}p$ data including the characteristics 
of kaonic hydrogen \cite{2012CS}. 

Concerning the the $P$-wave potential (\ref{eq:V1+}) we note that it contains no additional 
free parameter as the bare value of $\Sigma(1385)$ baryon mass is to be adjusted 
in a way that the position of the $\Sigma(1385)$ resonance is well reproduced. 
By doing so we got ${\tilde M}_{\Sigma^{*0}} = 1590$ MeV. 
The relatively large difference between the bare and the physical mass of the 
$\Sigma(1385)$ resonance is due to the fact that the pole shift is closely correlated to the
value of the inverse range parameter $\alpha$. Since it was our intention to have a fully fixed
P-wave model, we decided to use the value of $\alpha$ determined in the preceding
analysis. We prefer this approach to fixing either $\alpha$ or $\tilde{M}_{\Sigma^{*0}}$ at some arbitrary point
as well as to performing an additional fit of these values to the experimental data that we aim to predict.

\begin{figure}
\includegraphics[scale=0.25]{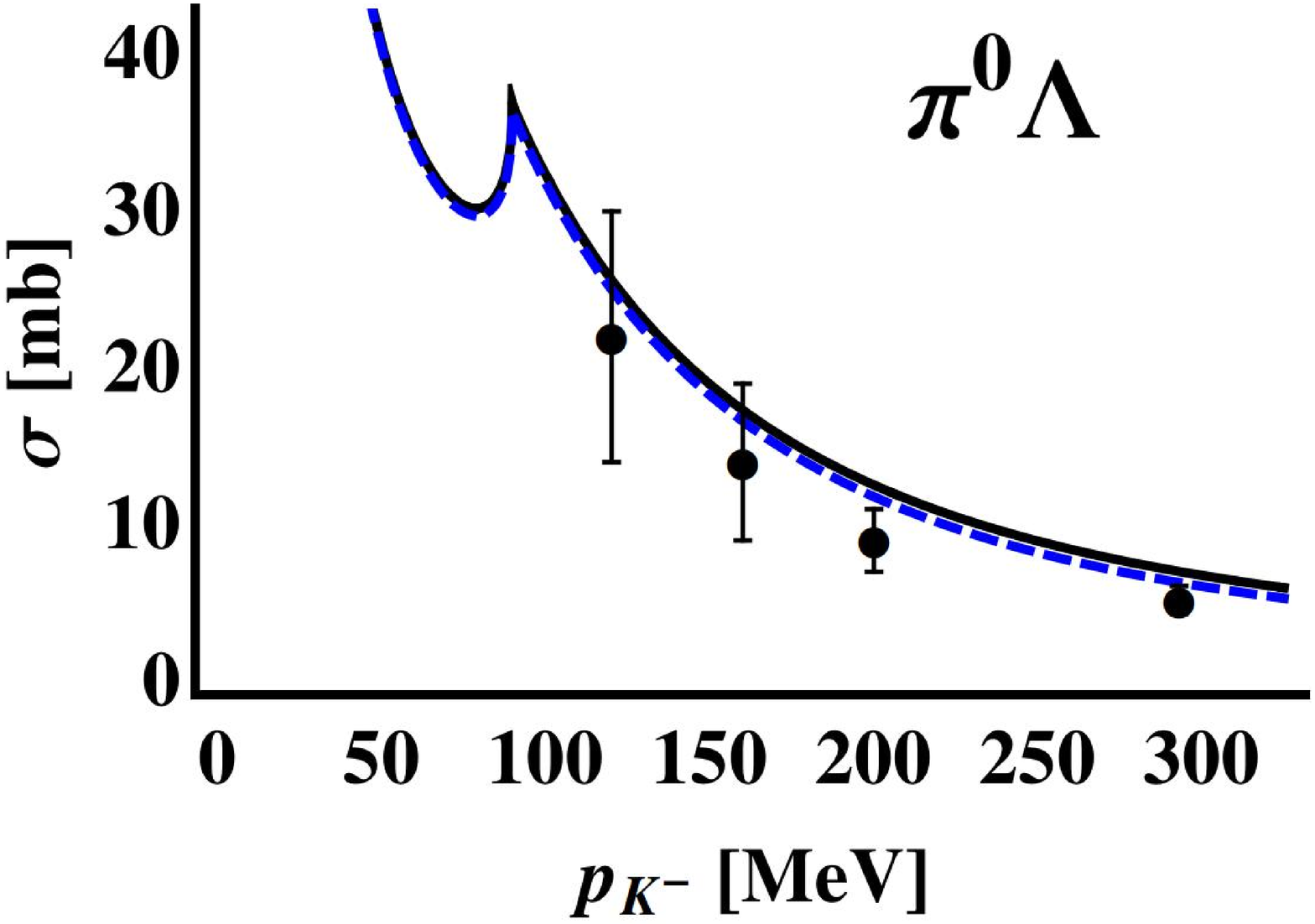}
\includegraphics[scale=0.25]{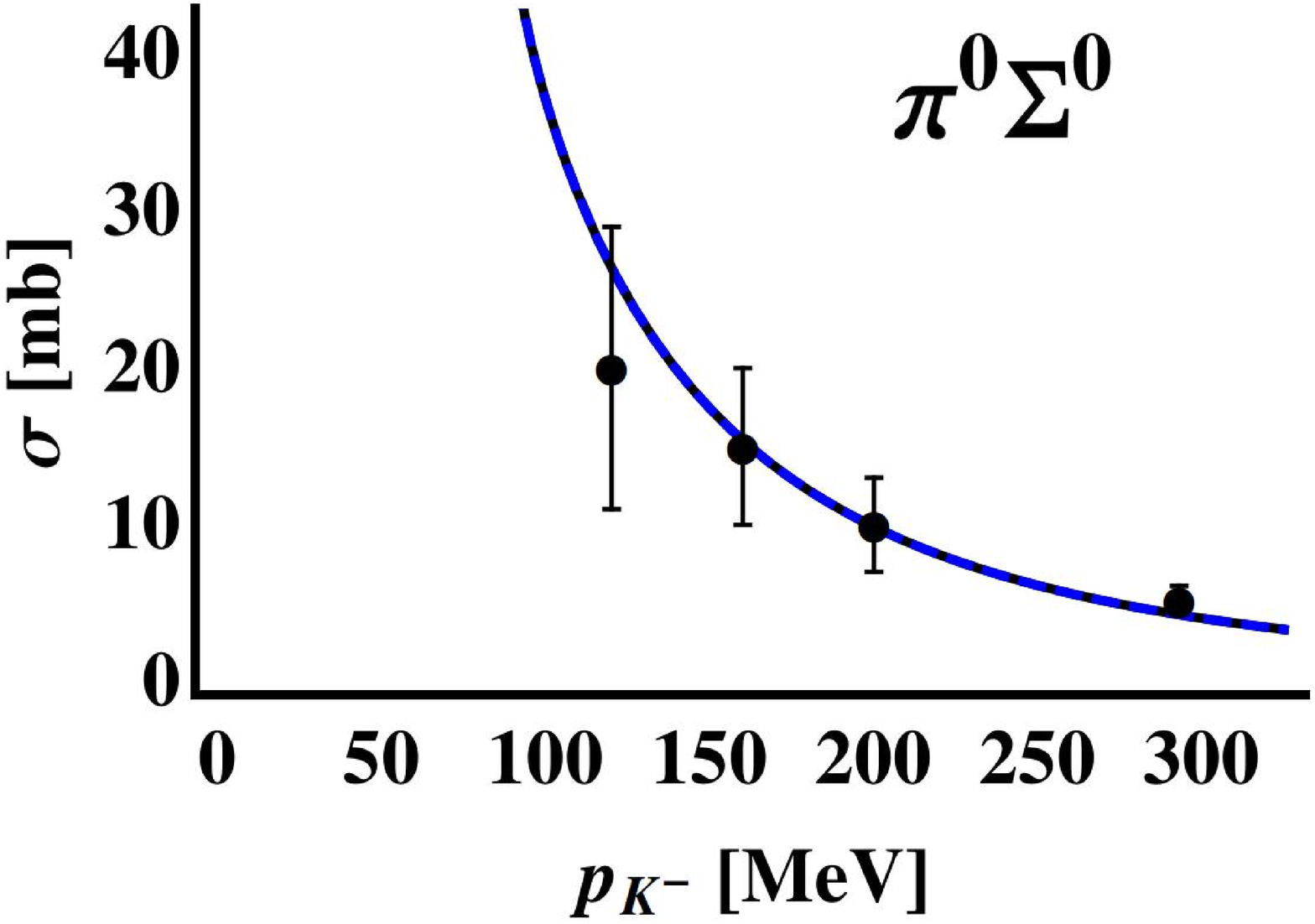} \\
\includegraphics[scale=0.25]{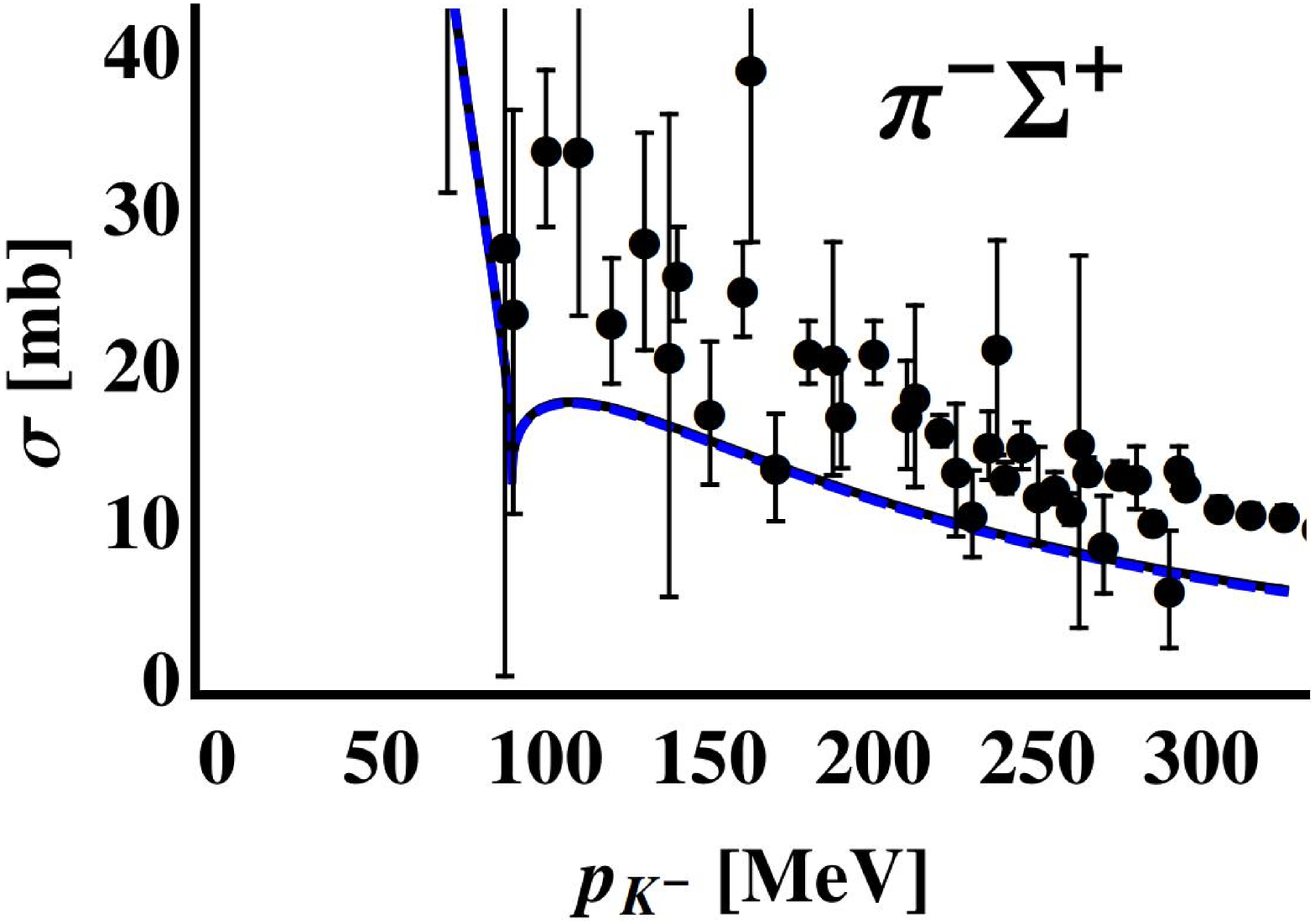}
\includegraphics[scale=0.25]{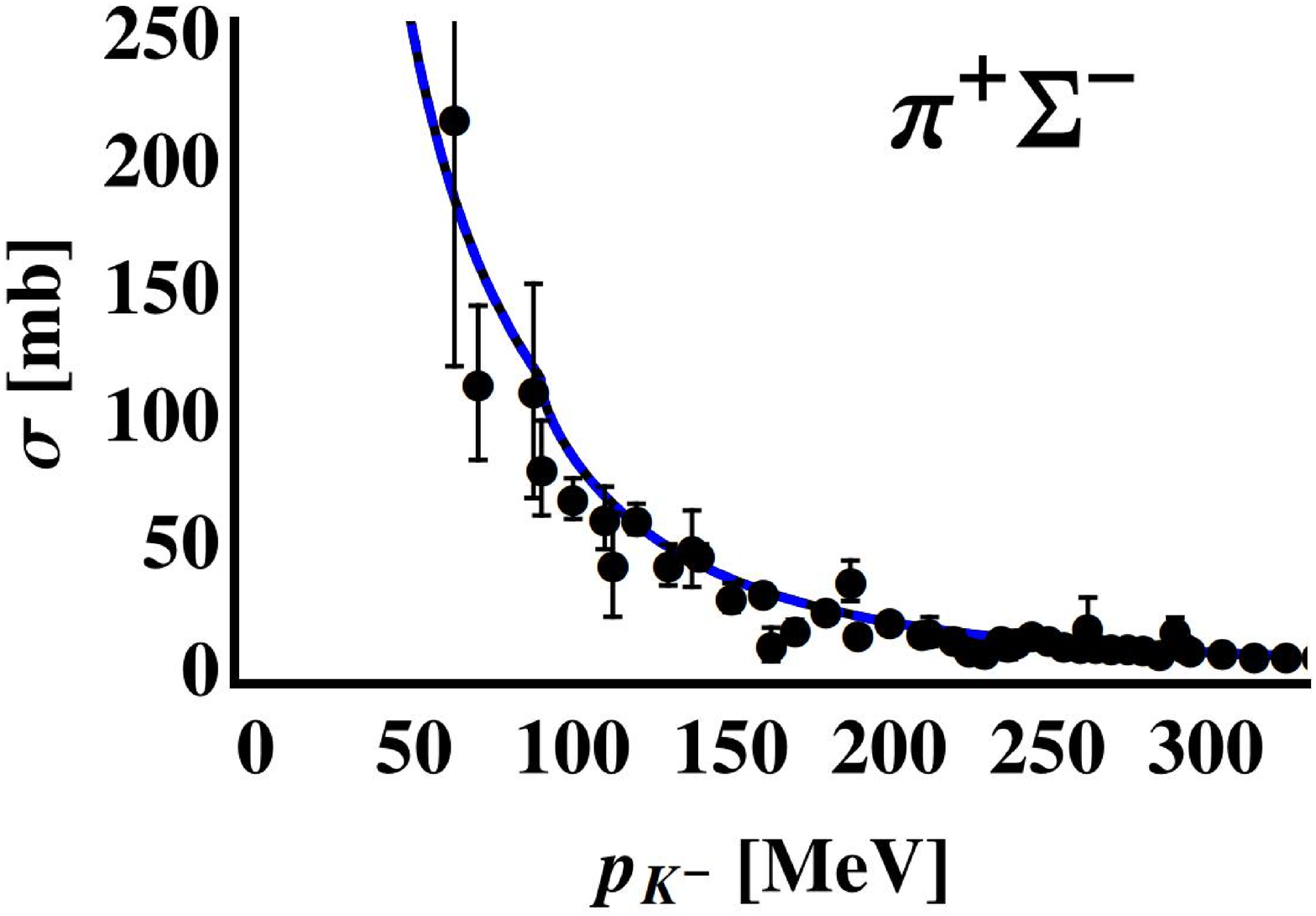} \\
\includegraphics[scale=0.25]{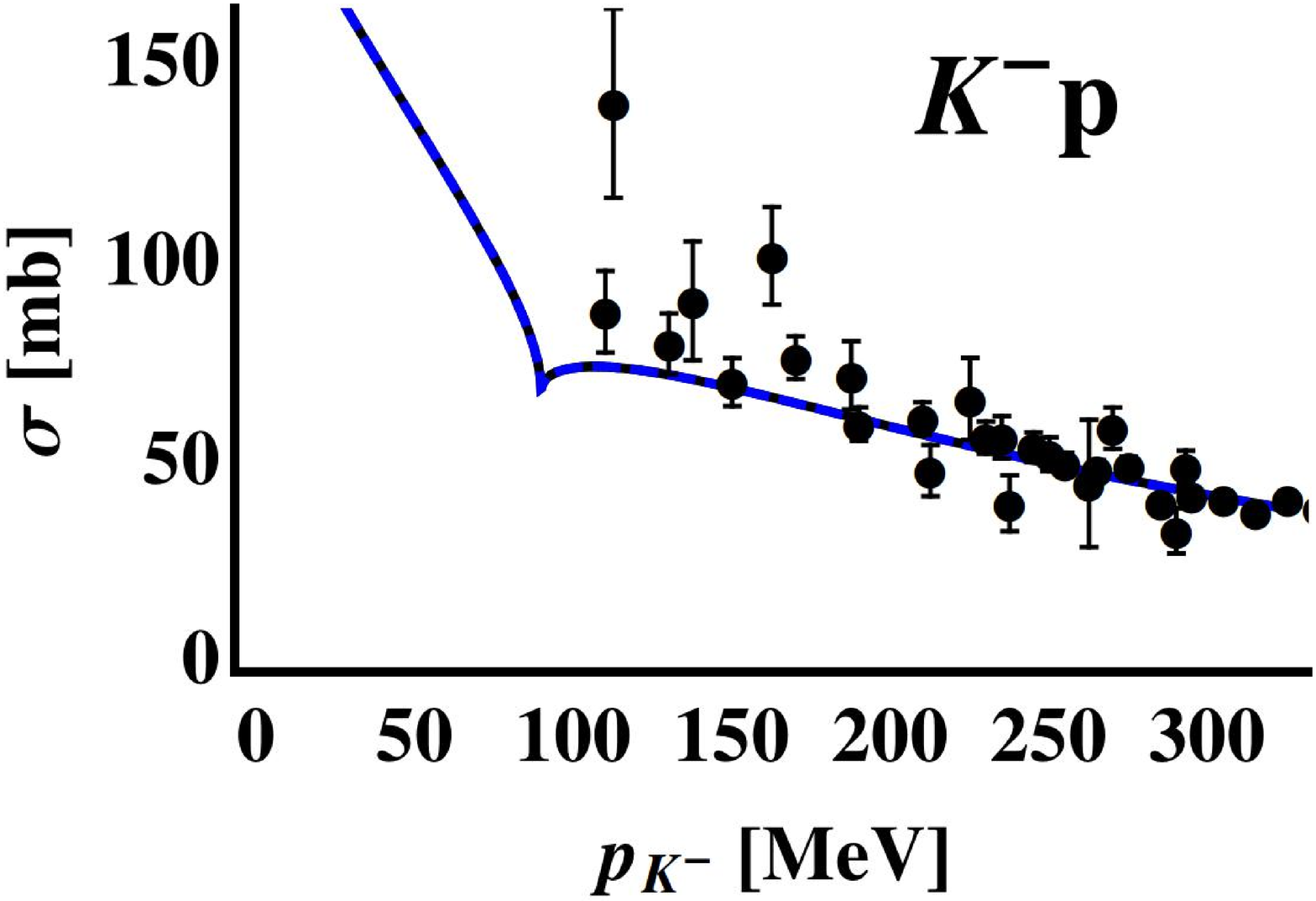} 
\includegraphics[scale=0.25]{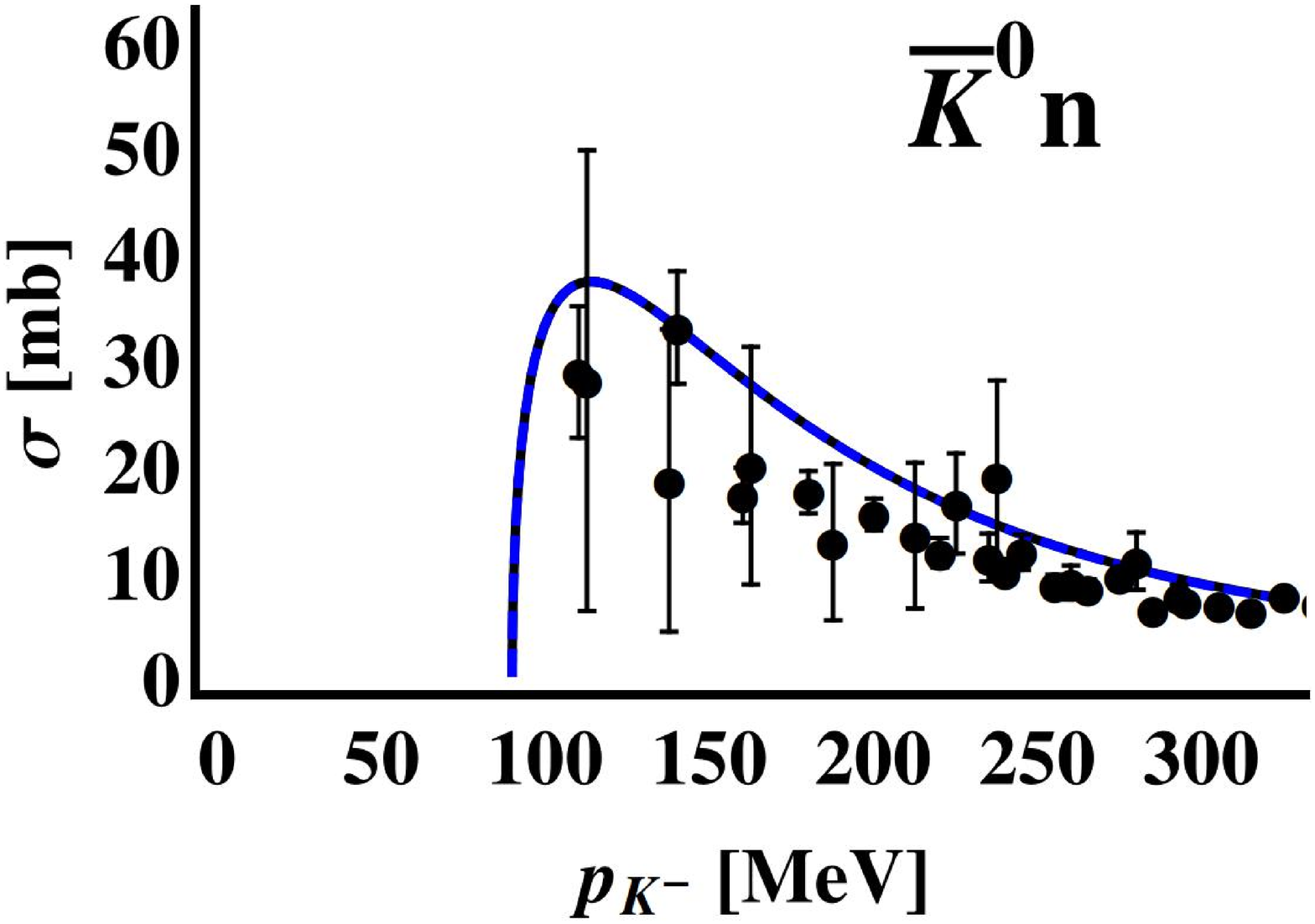}
\caption{Total $K^-p$ cross sections for $\pi^0 \Lambda$, $\pi^0 \Sigma^0$, $\pi^- \Sigma^+$, 
$\pi^+ \Sigma^-$, $K^- p$, and $\overline{K}^0 n$ channels (from the top left). 
Experimental data are from \cite{1982Cib}, \cite{1983Ev}, \cite{1965Sak}, \cite{1962HR}, 
\cite{1976Ma} \cite{1981Ban}, \cite{1963Wa}. Full line represents the full model 
(S and P waves) whereas the dashed line corresponds to the TW1 model (only S wave).}
\label{fig:totalCS}
\end{figure}

With all parameters of our effective anti-kaon nucleon model fixed we demonstrate its ability 
to reproduce the available $K^-p$ reactions data at low energies. The total cross sections 
for the $K^-p$ scattering into various channels are shown in Fig.~\ref{fig:totalCS}.
One can clearly see that the $S$-wave contribution dominates the total cross section 
in the whole energy region considered (up to 300 MeV kaon momentum). This observation 
is in agreement with a naive expectation that the $L=1$ partial wave contribution 
is small at energies considered in our analysis.

\begin{figure}
\includegraphics[scale=0.25]{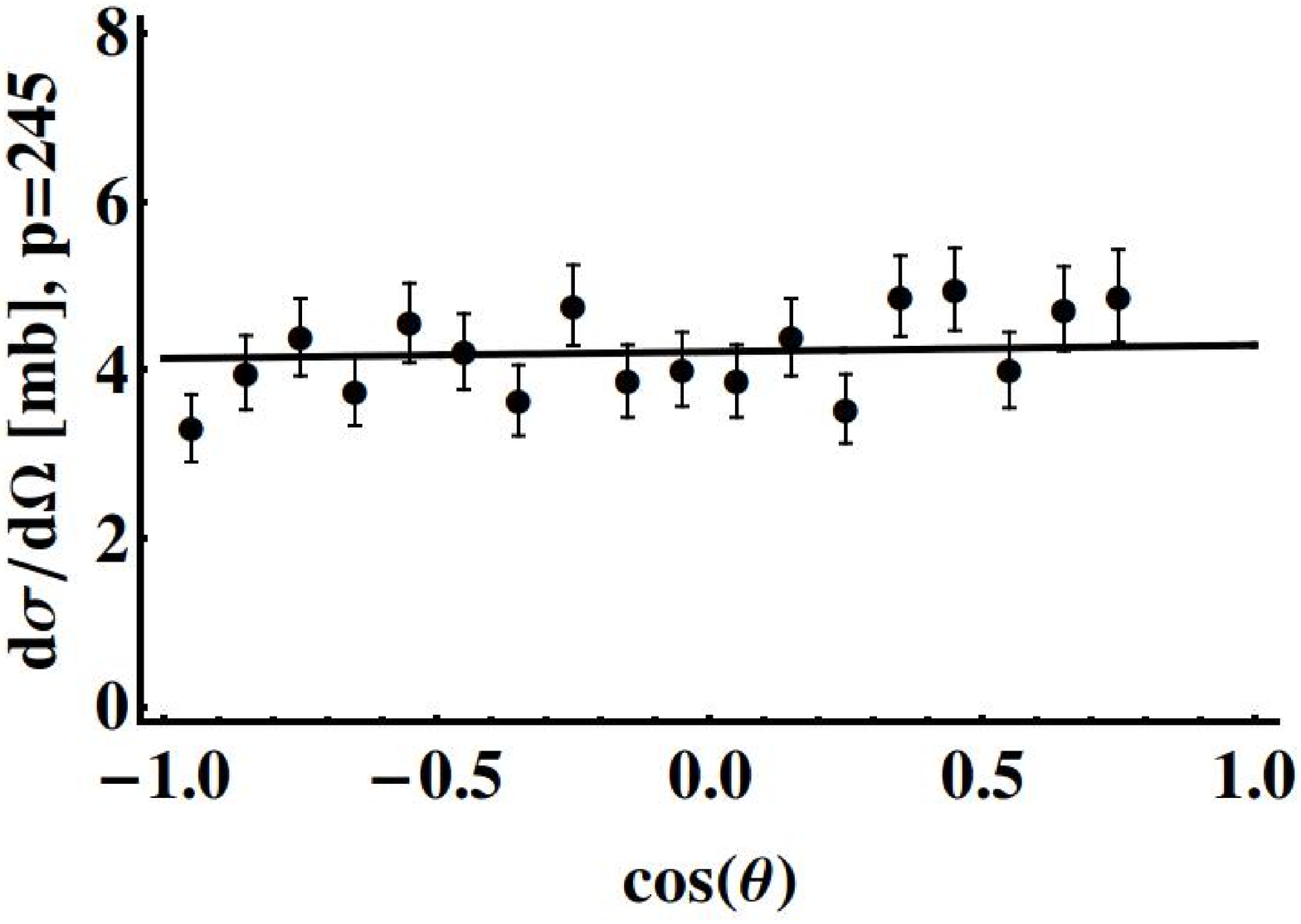}
\includegraphics[scale=0.25]{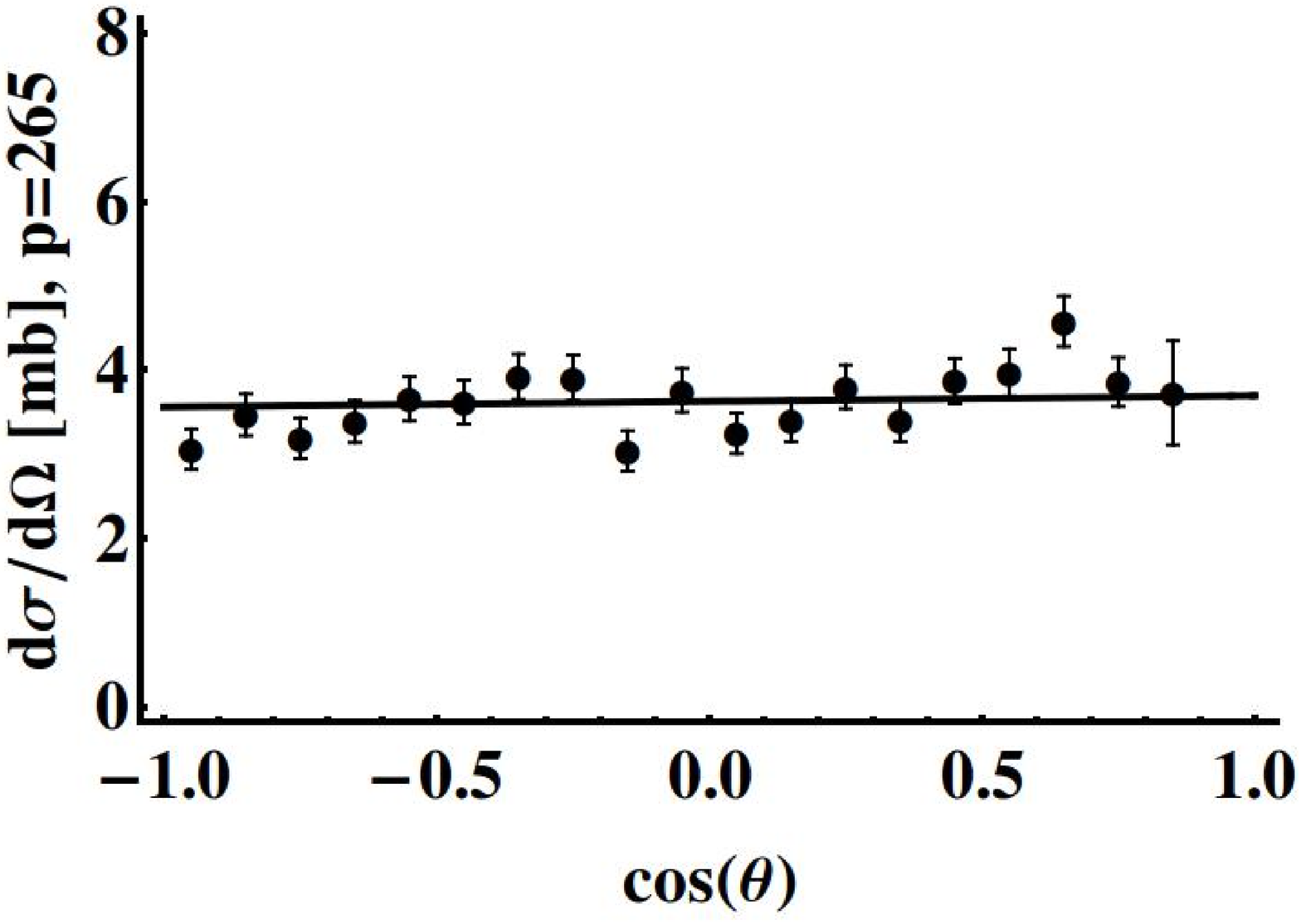}\\
\includegraphics[scale=0.25]{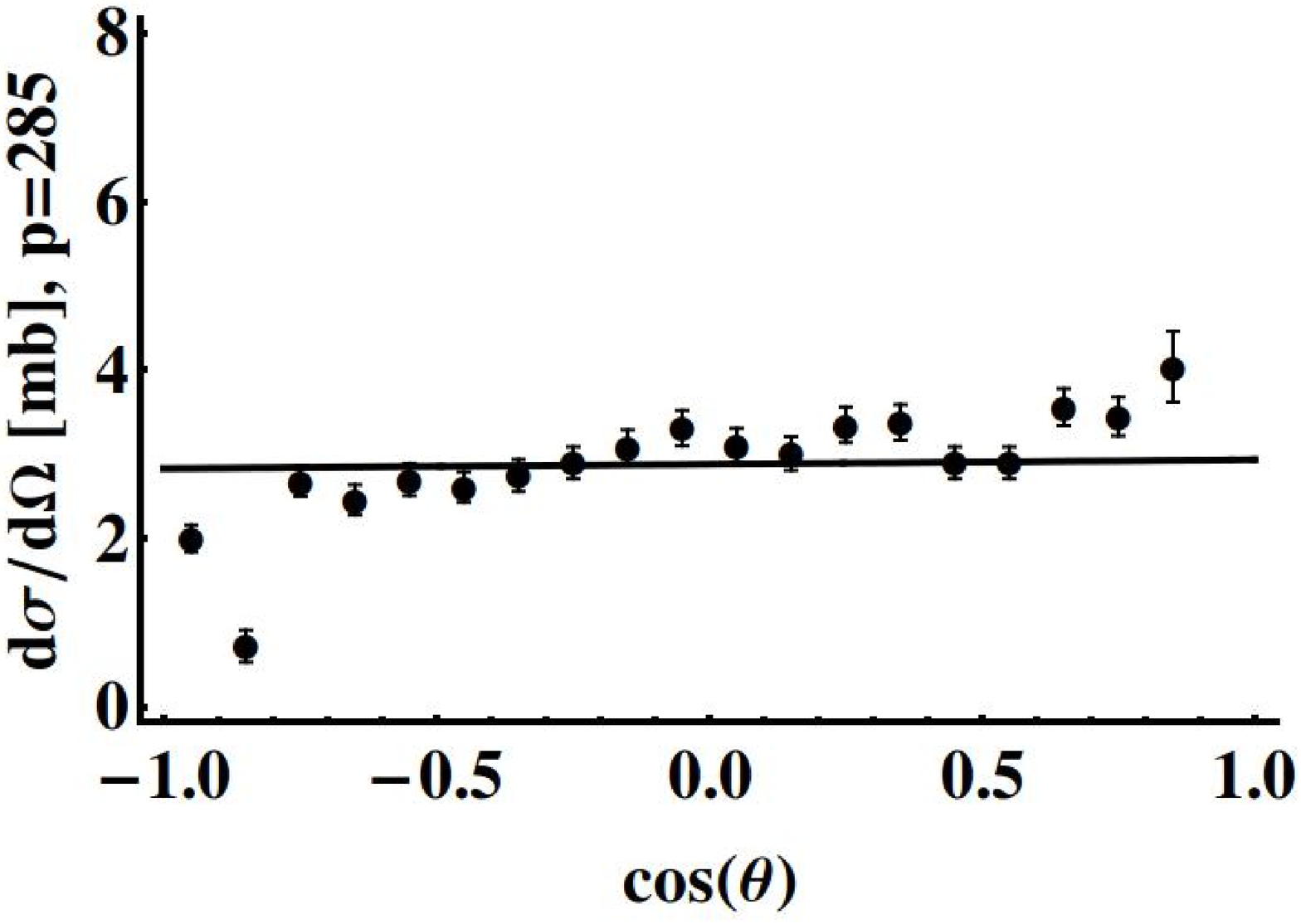}
\includegraphics[scale=0.25]{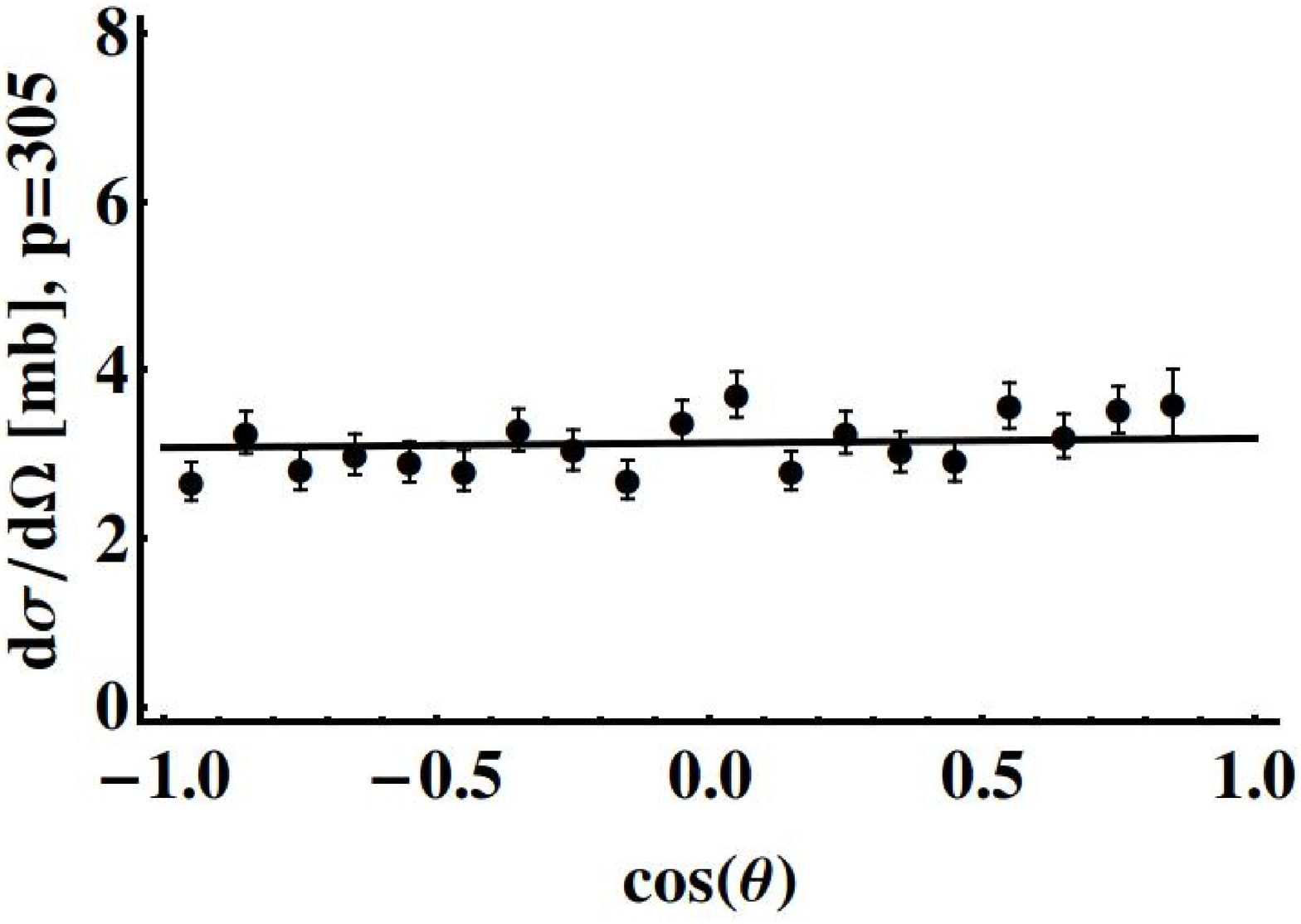}
\caption{Differential cross section for the $K^-p \rightarrow K^-p$ reaction at various energies.
Experimental data are from \cite{1976Ma}. Theoretical curves are normalized so that the total cross section is equal to the experimental one.}
\label{fig:K-pdifCS}
\end{figure}

\begin{figure}
\includegraphics[scale=0.25]{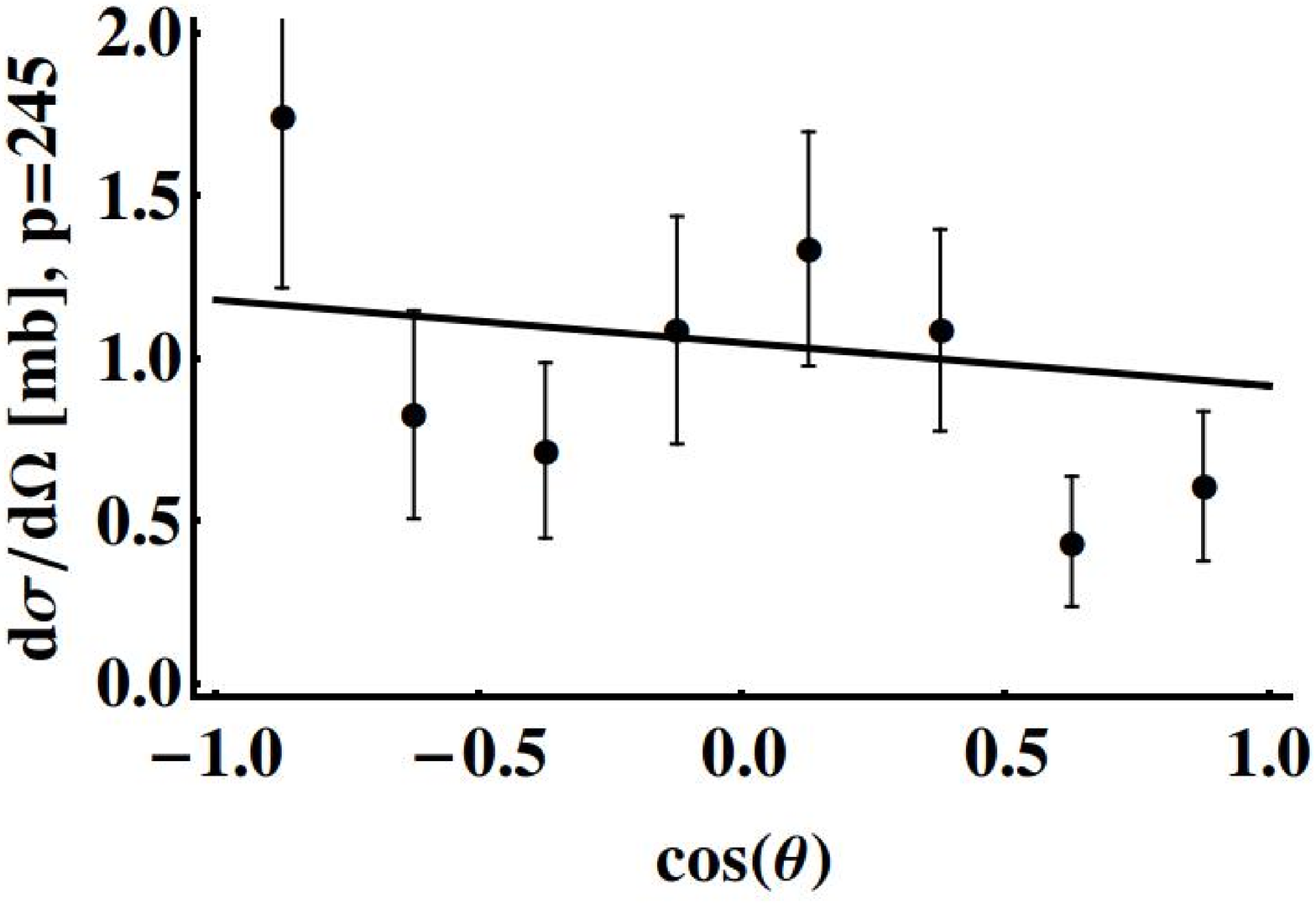}
\includegraphics[scale=0.25]{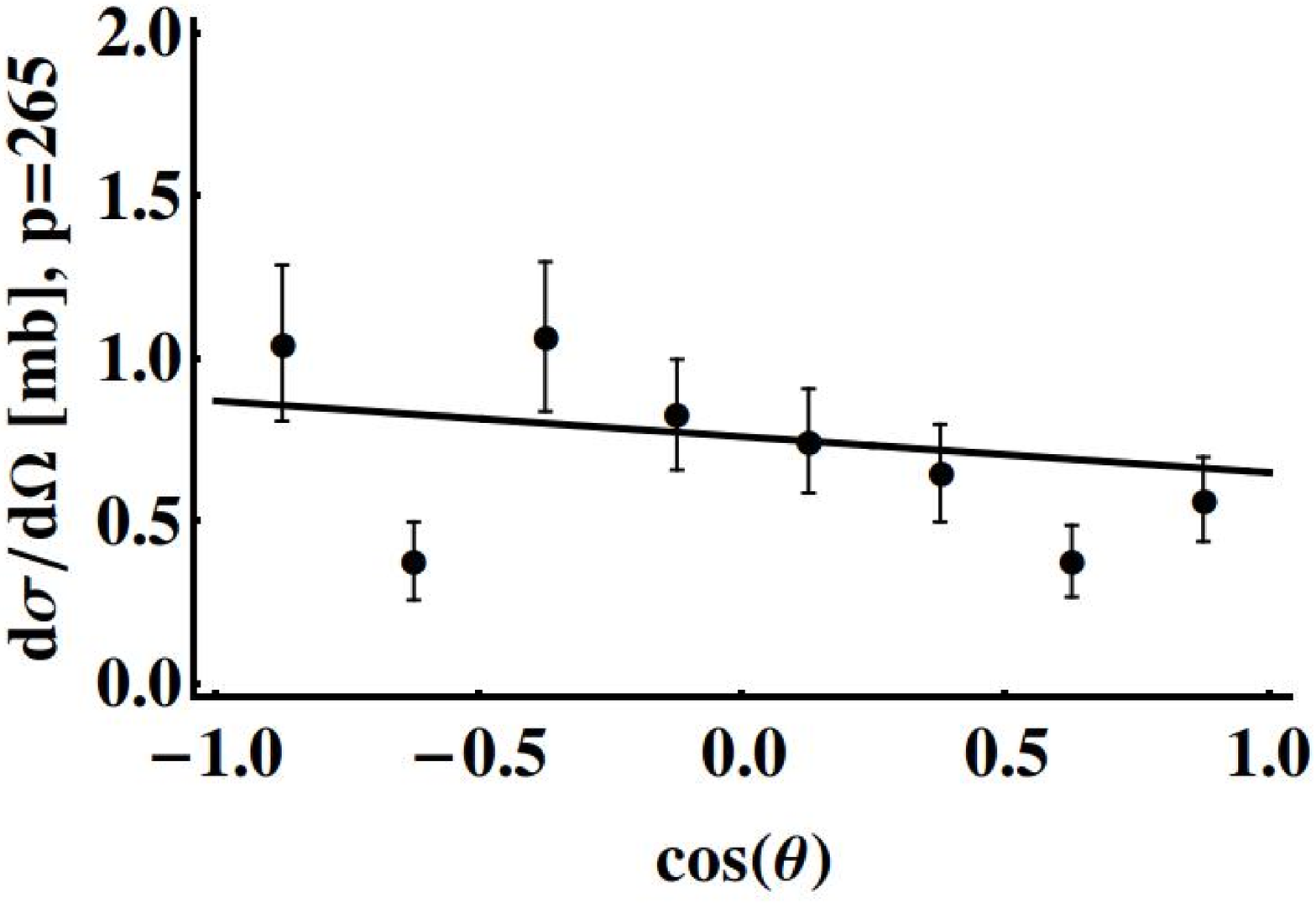}\\
\includegraphics[scale=0.25]{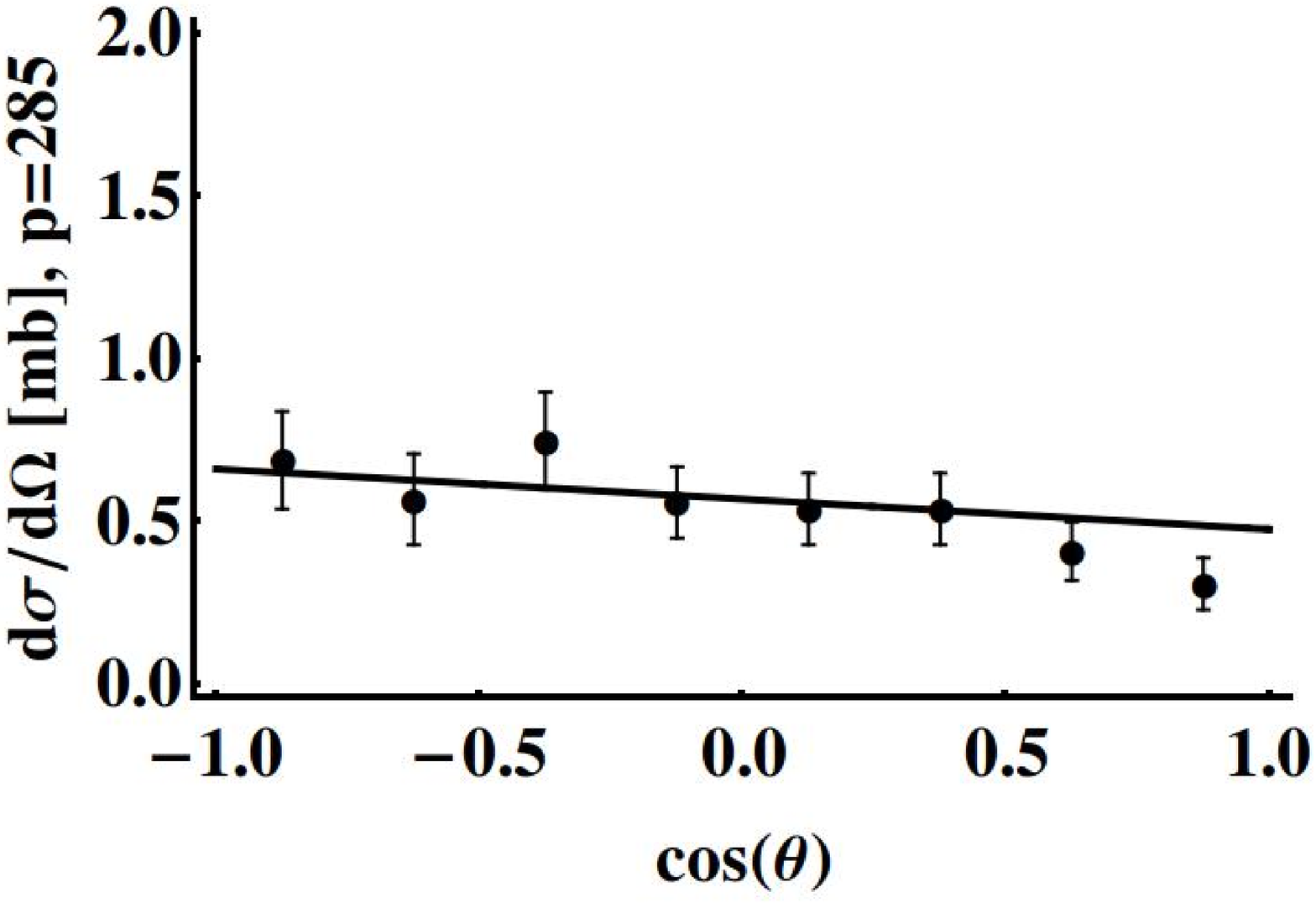}
\includegraphics[scale=0.25]{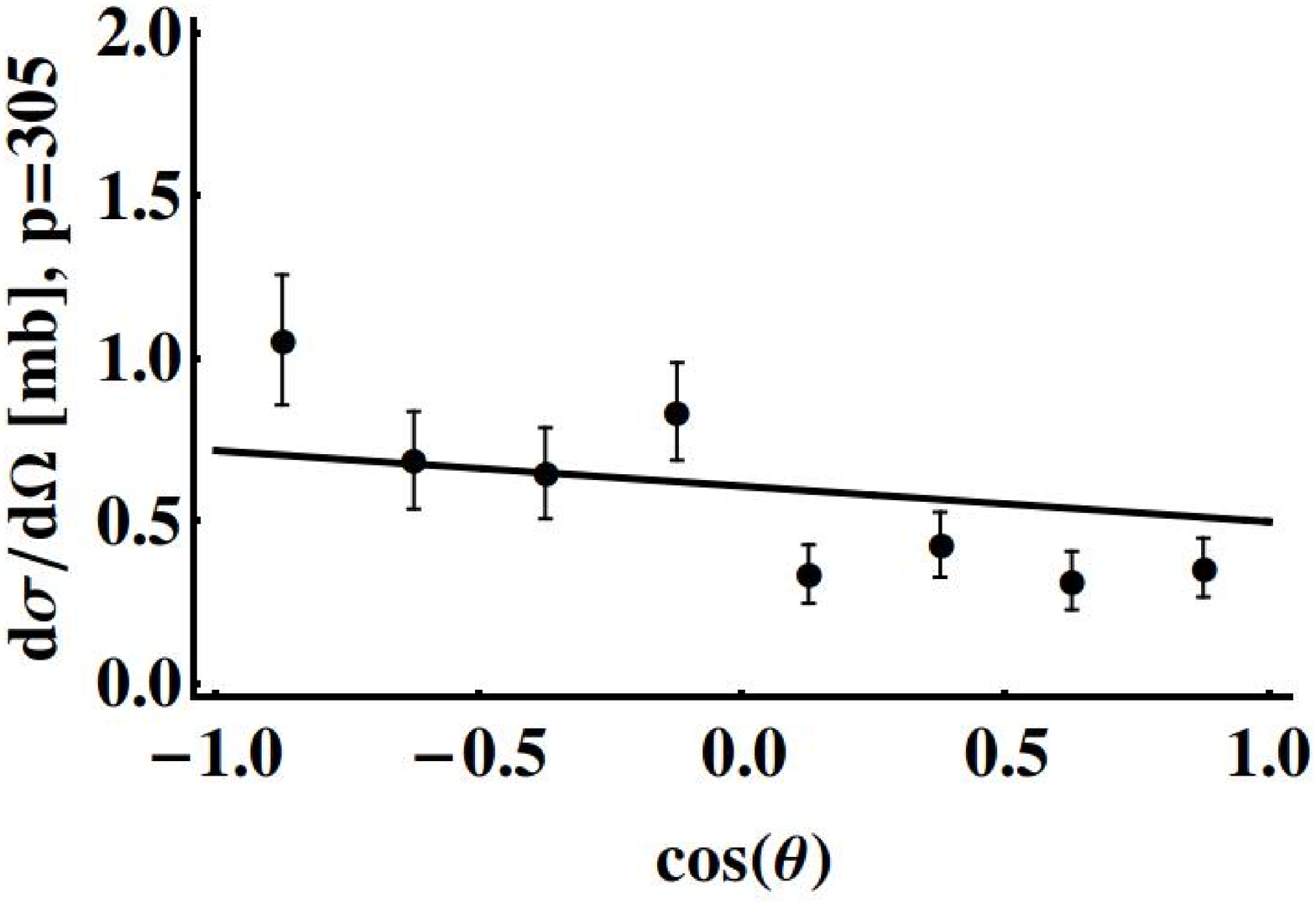}
\caption{Differential cross section for the $K^-p \rightarrow \bar{K}^0n$ reaction at various energies.
Experimental data are from \cite{1976Ma}. Theoretical curves are normalized 
so that the total cross section is equal to the experimental one.}
\label{fig:K0ndifCS}
\end{figure}

The physical quantity where $P$-wave physics plays a crucial role (and therefore can serve as an
honest testing ground for the present model) is the angular distribution of the scattering cross section.
In Figs.~\ref{fig:K-pdifCS} and \ref{fig:K0ndifCS}, the differential cross sections for the elastic
and charge-exchange reactions are shown. It is nice to observe that values predicted by our model are 
in a very good agreement with the experimental data. The direction of the slope 
(increasing for the elastic and decreasing for the charge-exchange reactions) 
is unquestionably reproduced. The angle of the slope is slightly underestimated 
for $K^-p \rightarrow K^-p$, whereas for the $K^-p \rightarrow \bar{K}^0 n$ reaction 
it matches the experimental observation rather accurately. We also note that the data on 
the differential cross sections are far more accurate and consistent than the asymmetries 
used in the preceding work \cite{2012Kre}. 

\begin{figure}[htb]
\includegraphics[scale=0.7]{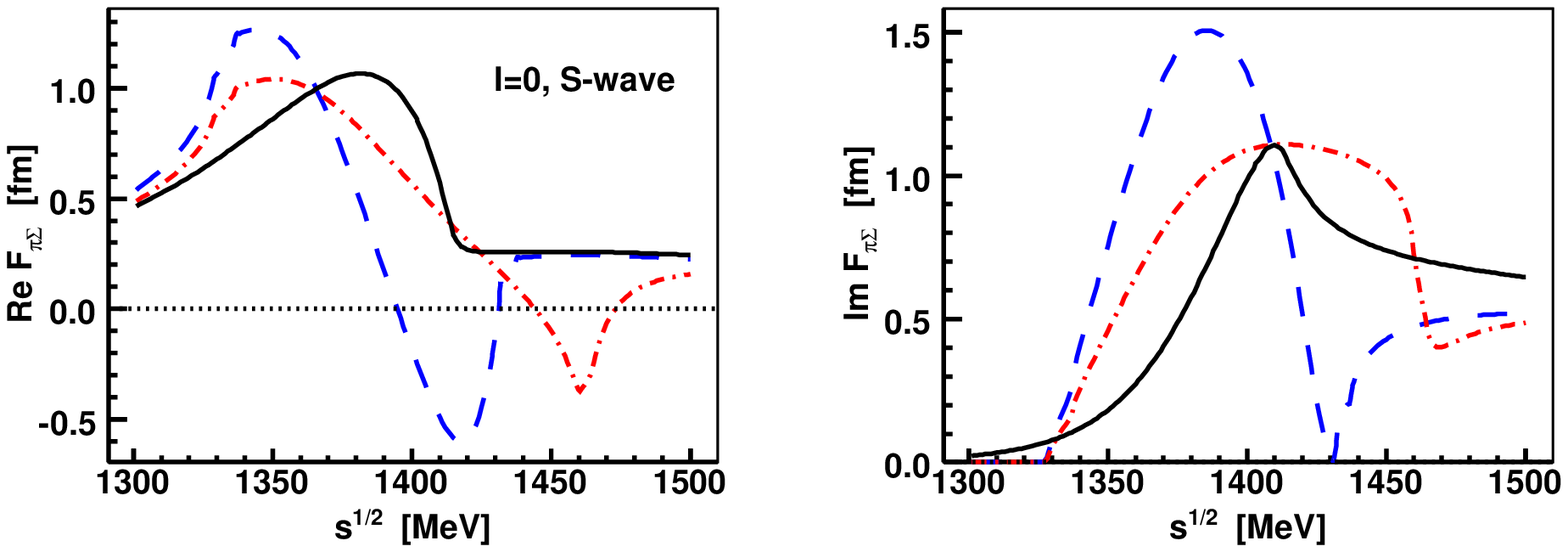}\\
\includegraphics[scale=0.7]{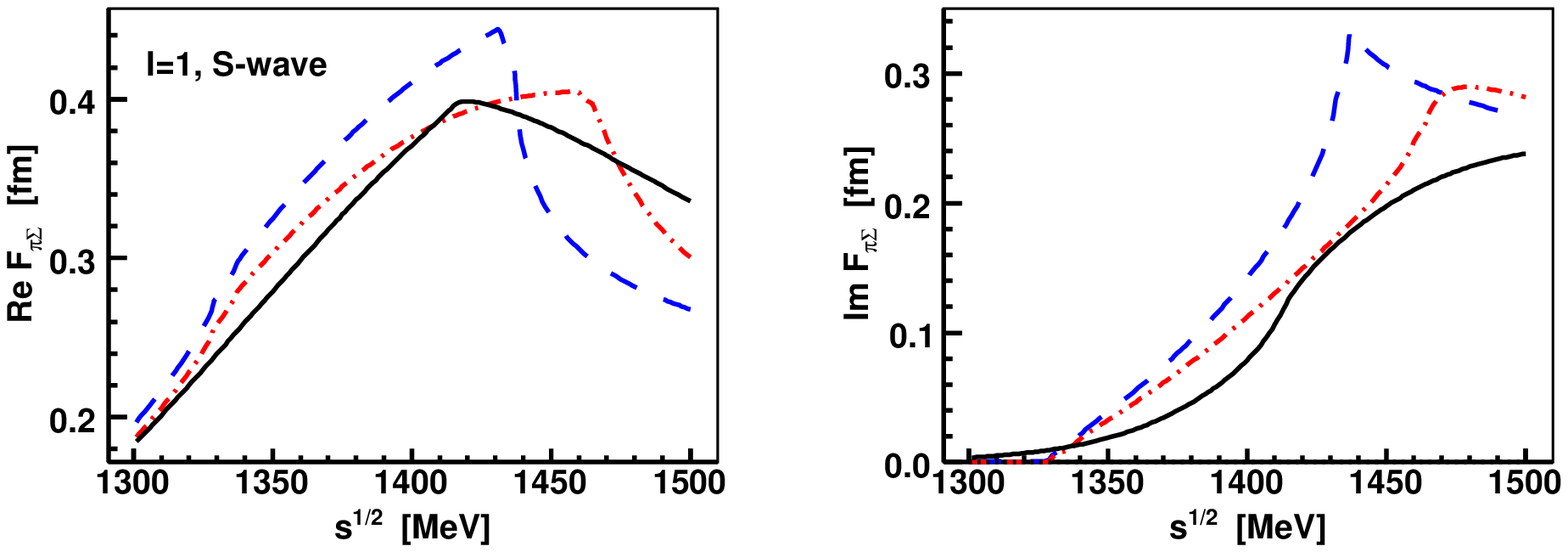}\\
\includegraphics[scale=0.7]{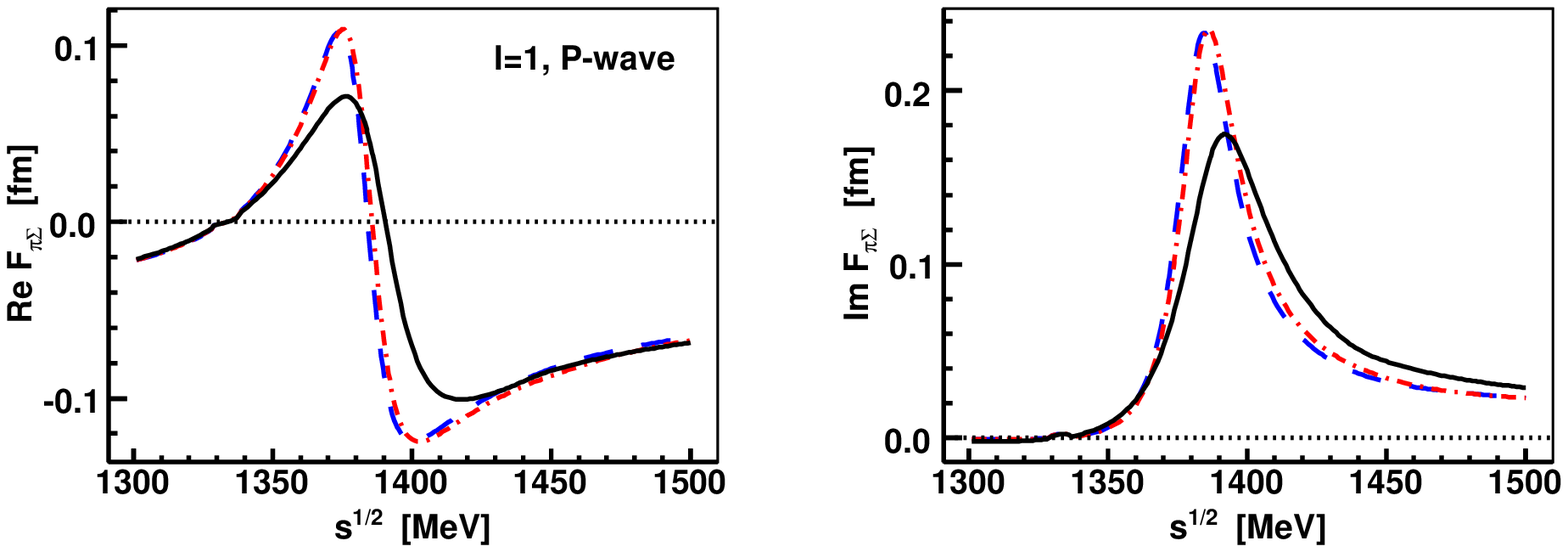}
\caption{Energy dependence of the $\pi\Sigma$ amplitudes. The dashed (blue) lines show the free space results, 
the dot-dashed (red) and continuous lines demonstrate the effects of including Pauli blocking only and Pauli blocking together with hadron selfenergies, respectively.}
\label{fig:FpiSig}
\end{figure}

Finally, with all the ingredients in place, we use the model to study impacts of nuclear matter 
on the shape of $\pi\Sigma$ mass distributions and on selected branching ratios that might 
be relevant for the analysis of the spectra measured experimentally. We start our analysis 
by demonstrating the nuclear medium effects on the $\pi\Sigma$ amplitudes. The amplitudes 
are shown in Fig.~\ref{fig:FpiSig} for the two discussed partial waves and isospins $I=0$ 
and $I=1$. The left and right panels present the real and imaginary parts of the amplitudes, 
respectively. The isoscalar $\pi\Sigma$ $P$-wave amplitudes are omitted in our analysis
as their impact on the discussed physical phenomena is negligible in the energy interval 
around the $\bar{K}N$ threshold. We also disregard the isotensor part of the $\pi\Sigma$
amplitude.

The $S$-wave part of the isoscalar $\pi\Sigma$ amplitude shown at the top panels of Fig.~\ref{fig:FpiSig} 
clearly demonstrates the importance of the $\Lambda(1405)$ resonance. At first, let us have a look at 
the lines that represent the free-space results. As the $\pi\Sigma$ channel couples more strongly 
to the lower of the two $I=0$ poles generated by the model the peak in the imaginary part 
of the free-space amplitude is located below 1400 MeV. However, since the $\pi\Sigma$-$\bar{K}N$ coupling is quite 
strong, the second pole (related to $\bar{K}N$ quasibound state) has also an impact on the amplitude 
and results in the second sign reversal of the real part of the amplitude. Both points at which the real part 
of the amplitude crosses the zero axis relate reasonably well to the positions of the two poles assigned 
to the $\Lambda(1405)$. For the TW1 model used in the present work the poles are located at complex energies
$z_{1} = (1371,-54)$ MeV and $z_{2} = (1433,-25)$ MeV. The effect of Pauli blocking imposed on the $\pi\Sigma$ 
system is visualized by the dot-dashed lines in the figure. The resonance structure is moved to higher energies 
and partially dissolves in the nuclear matter. 
The additional inclusion of hadron selfenergies shown by the continuous lines moves the structure 
back to lower energies, about 20-30 MeV below the $\bar{K}N$ threshold. Particularly, the nucleon self-energy 
is responsible for most of the shift with the other hadron self-energies having a moderate impact 
on the shape and width of the resonance. As far as the $\bar{K}N$ related pole assigned to 
the $\Lambda(1405)$ remains close to the real axis the resonance remains relatively narrow, 
interposed on the continuum formed by the second $\pi\Sigma$ related pole and by the other 
involved channels. These results are quite in line with observations made for the $\bar{K}N$ 
amplitudes in \cite{2012CS} and \cite{2011CFGGM}.

The $S$-wave part of the isovector $\pi\Sigma$ amplitude is given in the middle panels of Fig.~\ref{fig:FpiSig}.
Apparently, its magnitude is much smaller than the isoscalar one dominated by the presence of $\Lambda(1405)$. 
Still, a sharp peak at the $\bar{K}N$ threshold observed in the imaginary part of the amplitude indicates a presence 
of a pole nearby. Such isovector pole was already reported in \cite{2001OM} and its existence was argued for in 
Refs.~ \cite{2012CS} and \cite{2011CFGGM}. A possible existence of an isovector resonance close to 
the $\bar{K}N$ threshold may also be supported by the new CLAS data 
on the $\gamma p \longrightarrow K^{+}\pi^{\pm}\Sigma^{\mp}$ reactions \cite{2013CLAS}.

Finally, the bottom panels of Fig.~\ref{fig:FpiSig} present our results for the $P$-wave part of 
the isovector $\pi\Sigma$ amplitude. The imaginary part of the amplitude shows a resonance with 
a width of about $30$ MeV, in a reasonable agreement with the $\Sigma(1385)$ resonance width 
of $36$ MeV listed by the Particle Data Group \cite{2014PDG}. 
Unlike the $S$-wave amplitudes the Pauli blocking has little 
impact on the $P$-wave ones. We checked that the same applies to the $P$-wave part of the elastic 
$\bar{K}N$ amplitude, so the observation does not apply exclusively to the $\pi\Sigma$ channel. 
The feature is most likely related to our treatment of the $P$-wave interaction that is restricted 
to a formation of an intermediate $\Sigma(1385)$ resonance, see Eq.~(\ref{eq:Vdir10}). 
The inclusion of hadron self-energies practically does not affect the position of the resonance 
structure impacting only on the magnitude of the observed peak.
Apparently, the leading resonant interaction in the Born term is much stronger than higher order 
contributions of the expansion that include intermediate meson-baryon states affected 
by the nuclear medium. 

\begin{figure}[htb]
\includegraphics[scale=0.7]{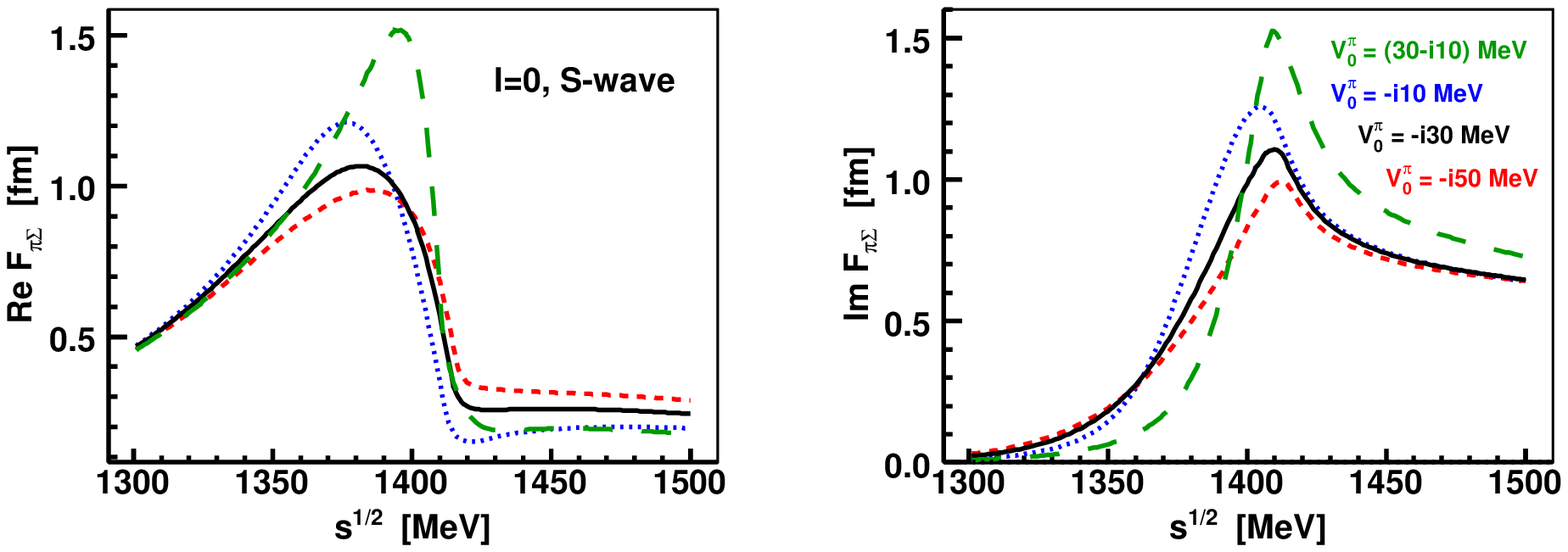}\\
\includegraphics[scale=0.7]{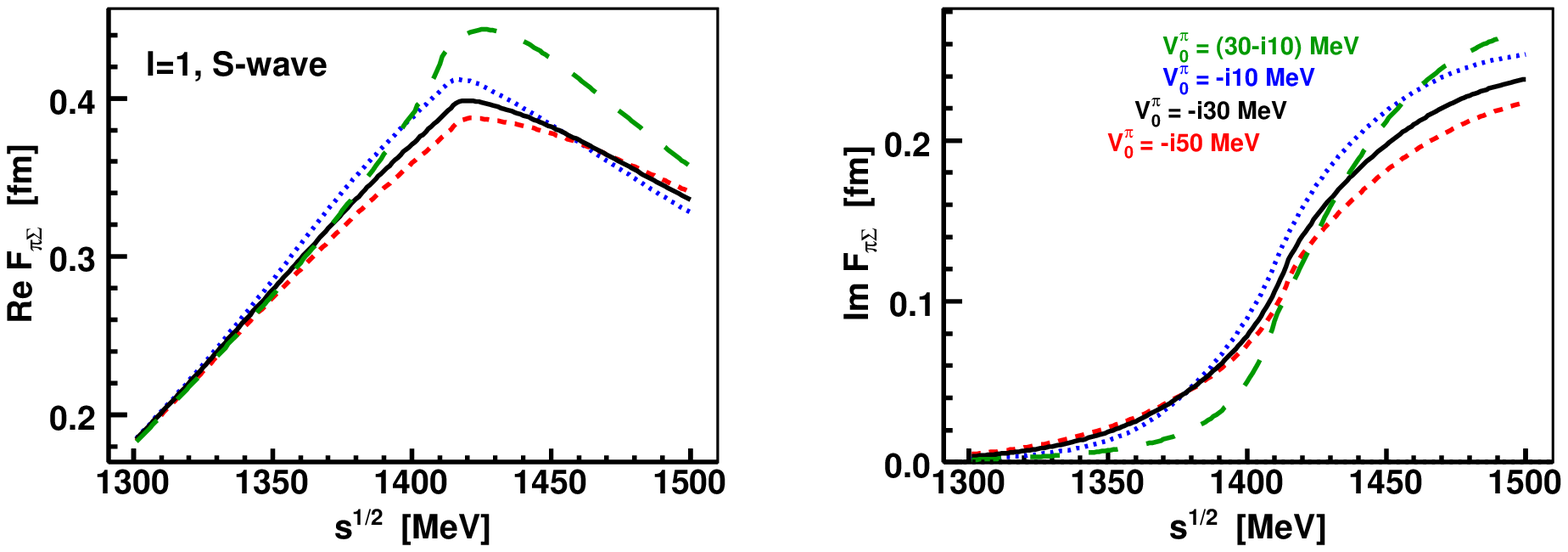}\\
\includegraphics[scale=0.7]{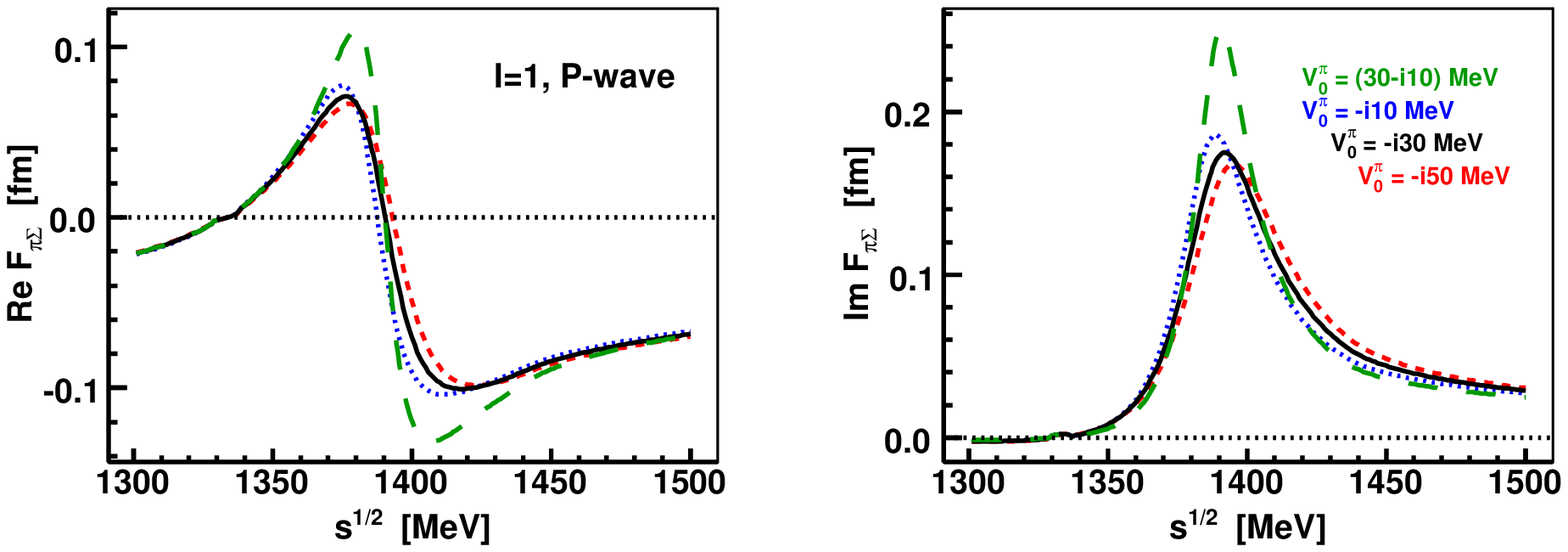}
\caption{A demonstration of the sensitivity of the in-medium $\pi \Sigma$ amplitudes to a magnitude 
of pion absorption in nuclear matter. The dotted (blue), continuous and short-dashed (red) lines were 
calculated with pion-nuclear optical potential depths of ${\rm i}V^{\pi}_0 = 10,\; 30$ and $50$ MeV, 
respectively. For a reference the dashed (green) lines show results obtained with a pion potential 
adopted in Ref.~\cite{2011CFGGM}.}
\label{fig:vpi}
\end{figure}

At this point we find it appropriate to mention the sensitivity of our in-medium results 
to the adopted form of the pion self-energy. The whole effect is demonstrated in Fig.~\ref{fig:vpi} 
for the same amplitudes as those in Fig.~\ref{fig:FpiSig}. The calculations were made for three 
different values of the purely imaginary pion-nuclear optical potential depth, i$V^{\pi}_{0} = 10$, 
$30$ and $50$ MeV, and for the optical potential depth of $V^{\pi}_{0} = (30 - {\rm i}10)$ MeV 
used previously in Ref.~\cite{2011CFGGM}. As observed in the figure, a feasible increase 
of the imaginary part $\Im V^{\pi}_{0}$ decreases the magnitude of the $\Lambda(1405)$ resonance 
structure shown in the top panels. Similarly, a smaller pion-nuclear absorption leads 
to a more pronounced in-medium resonance. The same, though much smaller, effects are seen in 
the bottom panels where the $\Sigma(1385)$ resonance structure is manifested. Interestingly, 
an addition of a repulsive nonzero real part to the pion-nuclear potential (as anticipated 
in \cite{2011CFGGM}) does not have a significant impact on the position of the $\Lambda(1405)$ 
and $\Sigma(1385)$ structures, but the magnitude of both peaks gets increased. As there is 
no genuine peak in the in-medium isovector $\pi \Sigma$ amplitude shown in the middle panels 
of the figure the variation of pion self-energy has a relatively small impact there. 
The results shown in Fig.~\ref{fig:vpi} can also be viewed as a demonstration of theoretical 
ambiguities and a measure of anticipated theoretical errors in our predictions. We also note that 
our results obtained for the largest pion absorption, i$V^{\pi}_{0} = 50$, MeV are in a reasonable  
agreement with observations made in \cite{2006TRO} where a more sophisticated formulation 
of the pion-nuclear optical potential was adopted. On the other hand the authors of \cite{2006TRO} 
consider an unlikely $\Sigma$-nuclear attractive interaction in their model and it remains 
to be seen whether their treatment of in-medium pions based on particle-hole and $\Delta$-hole 
excitations is really a realistic one.

To complete the analysis of theoretical ambiguities we also checked the sensitivity 
of our results to the variations of (not so well established) imaginary parts of the 
baryon self-energies. We have found that neglecting completely the imaginary parts of 
the nucleon and $\Lambda$ self-energies, or increasing the imaginary part of the $\Sigma$ 
optical potential depth to $\Im V^{\Sigma}_{0} = -{\rm i}20$ MeV has marginal effect on 
the computed in-medium $\pi \Sigma$ amplitudes.

\begin{figure}[htb]
\includegraphics[scale=0.5]{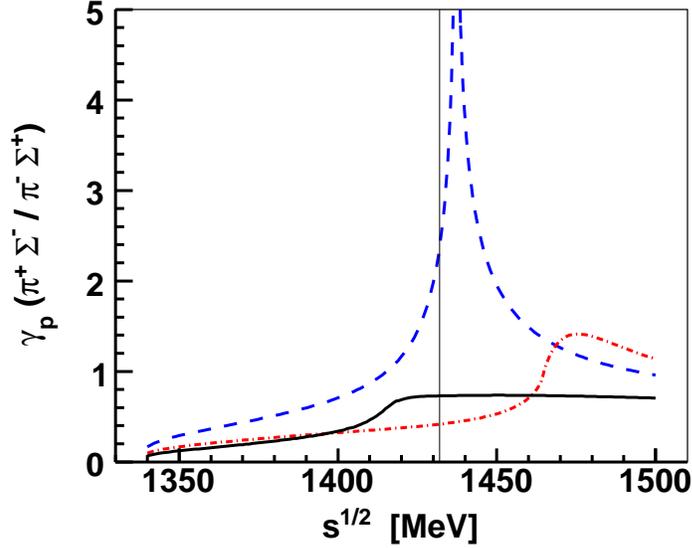}
\caption{Energy dependence of the branching ratio $\gamma_p = \Gamma(\pi^{+}\Sigma^{-})/\Gamma(\pi^{-}\Sigma^{+})$. 
The dashed (blue) line shows the free space results, the dot-dashed (red) and continuous lines demonstrate 
the effects of including Pauli blocking only and Pauli blocking together with hadron selfenergies, respectively. 
The thin vertical line marks the $K^{-}p$ threshold.}
\label{fig:BR1}
\end{figure}

We now turn our attention to the energy and density dependence of the branching ratio 
\beq
\gamma_p = \frac{\sigma(K^{-}p \rightarrow \pi^{+}\Sigma^{-})}{\sigma(K^{-}p \rightarrow \pi^{-}\Sigma^{+})} 
= \frac{\Gamma(\pi^{+}\Sigma^{-})}{\Gamma(\pi^{-}\Sigma^{+})}  
\;,
\eeq{eq:BRp}
where the expression utilizing the transition probabilities $\Gamma(j) = 8\pi\mid F_{K^{-}p,\,j} \mid ^{2}$ 
can be used even for energies below the $K^{-}p$ threshold. The threshold value $\gamma_p = 2.360 \pm 0.040$ 
\cite{1981Mar} represents one of the experimental quantities fitted by the TW1 model. 
In Fig.~\ref{fig:BR1} we show its energy dependence and how it is affected 
by the nuclear medium due to the Pauli blocking and hadron selfenergies when evaluating 
the intermediate state meson-baryon propagator.

First of all it is worth mentioning the narrow resonant structure above the $K^{-}p$ threshold we get in the free space. 
The position of the peak coincides with an opening of the $\bar{K}^{0}n$ channel which is also marked by a sharp 
dip observed in Fig.~\ref{fig:totalCS} for the $\sigma(K^-p\rightarrow\pi^{-}\Sigma^{+})$ cross section. 
The $\gamma_p$ peak is related to a difference in decomposition of the two charged $\pi\Sigma$ states into parts with a 
specific isospin,
\beqa
\mid \pi^{-}\Sigma^{+} \rangle & = & -\sqrt{\frac{1}{3}} \mid \pi\Sigma \rangle_{I=0}  
                                     +\sqrt{\frac{1}{2}} \mid \pi\Sigma \rangle_{I=1} 
                                     -\sqrt{\frac{1}{6}} \mid \pi\Sigma \rangle_{I=2} \\
\mid \pi^{+}\Sigma^{-} \rangle & = & -\sqrt{\frac{1}{3}} \mid \pi\Sigma \rangle_{I=0} 
                                     -\sqrt{\frac{1}{2}} \mid \pi\Sigma \rangle_{I=1} 
                                     -\sqrt{\frac{1}{6}} \mid \pi\Sigma \rangle_{I=2} \; .
\eeqa{eq:PiSig}
Only the isoscalar and isovector parts contribute to reactions with $K^{-}p$ in the initial 
state and the discussed branching ratio can be written as
\beq
\gamma_p = \frac{\left| T_0 + \sqrt{\frac{3}{2}} T_1 \right|^{2}}
                {\left| T_0 - \sqrt{\frac{3}{2}} T_1 \right|^{2}} \; ,
\eeq{eq:BRiso}
where we introduced the isoscalar and isovector transition amplitudes 
$T_{I=0,1} = \langle \pi\Sigma \mid T \mid \bar{K}N \rangle_{I=0,1}$. 
The strong energy dependence of $T_0$ (an partly of the $T_1$ amplitude too) close to the $\bar{K}N$ 
threshold then leads to a narrow peak at energies where the amplitudes $T_0$ and $T_1$ combine 
in such a way that the denominator of Eq.~(\ref{eq:BRiso}) becomes small, i.e.~when it holds
\beq
\left|  T_0 - \sqrt{\frac{3}{2}} T_1 \right| << 
\left| T_0 + \sqrt{\frac{3}{2}} T_1 \right|  \; .
\eeq{}
In nuclear matter this condition is not satisfied, so the peak gets dissolved. Though, one still 
observes a shift of the structure related to $\Lambda$(1405) dynamics when the Pauli blocking 
and hadron selfenergies are accounted for.

The strong energy dependence of the $\gamma_p$ ratio makes its experimental value at the $\bar{K}N$ threshold 
quite useful in fixing the parameters of any model that aims at a reliable qualitative description of 
meson-baryon interactions in that energy region. We also mention that a similar branching ratio 
for processes on neutron, $\gamma_n = \sigma(K^{-}n \rightarrow \pi^{0}\Sigma^{-})/\sigma(K^{-}n \rightarrow \pi^{-}\Sigma^{0})$, 
is not so interesting as both cross sections are composed of only the isovector transition amplitudes, 
so there is no $T_0$ versus $T_1$ interference and the rate is compatible with one in the whole 
energy region.

\begin{figure}[htb]
\includegraphics[scale=0.4]{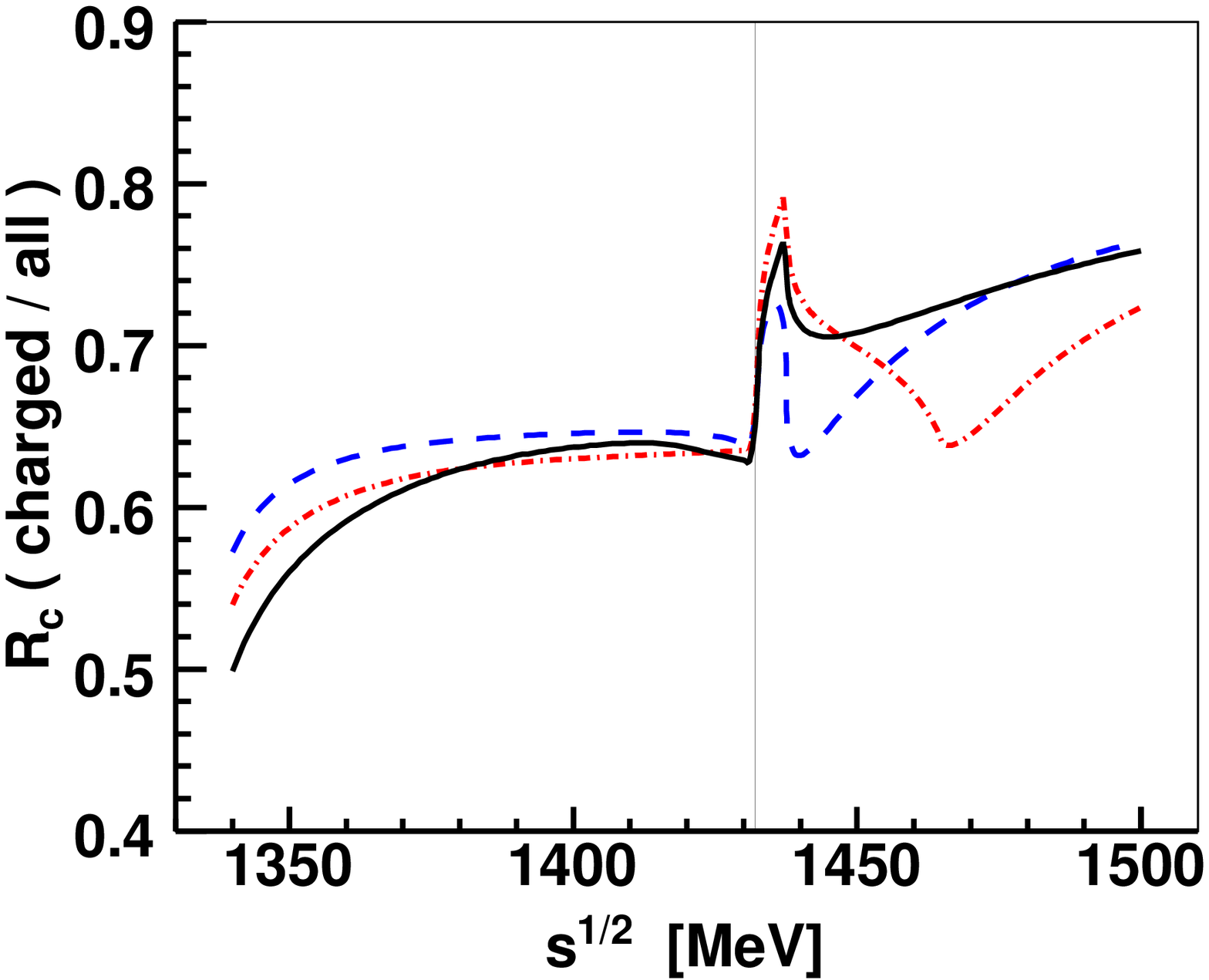}
\includegraphics[scale=0.4]{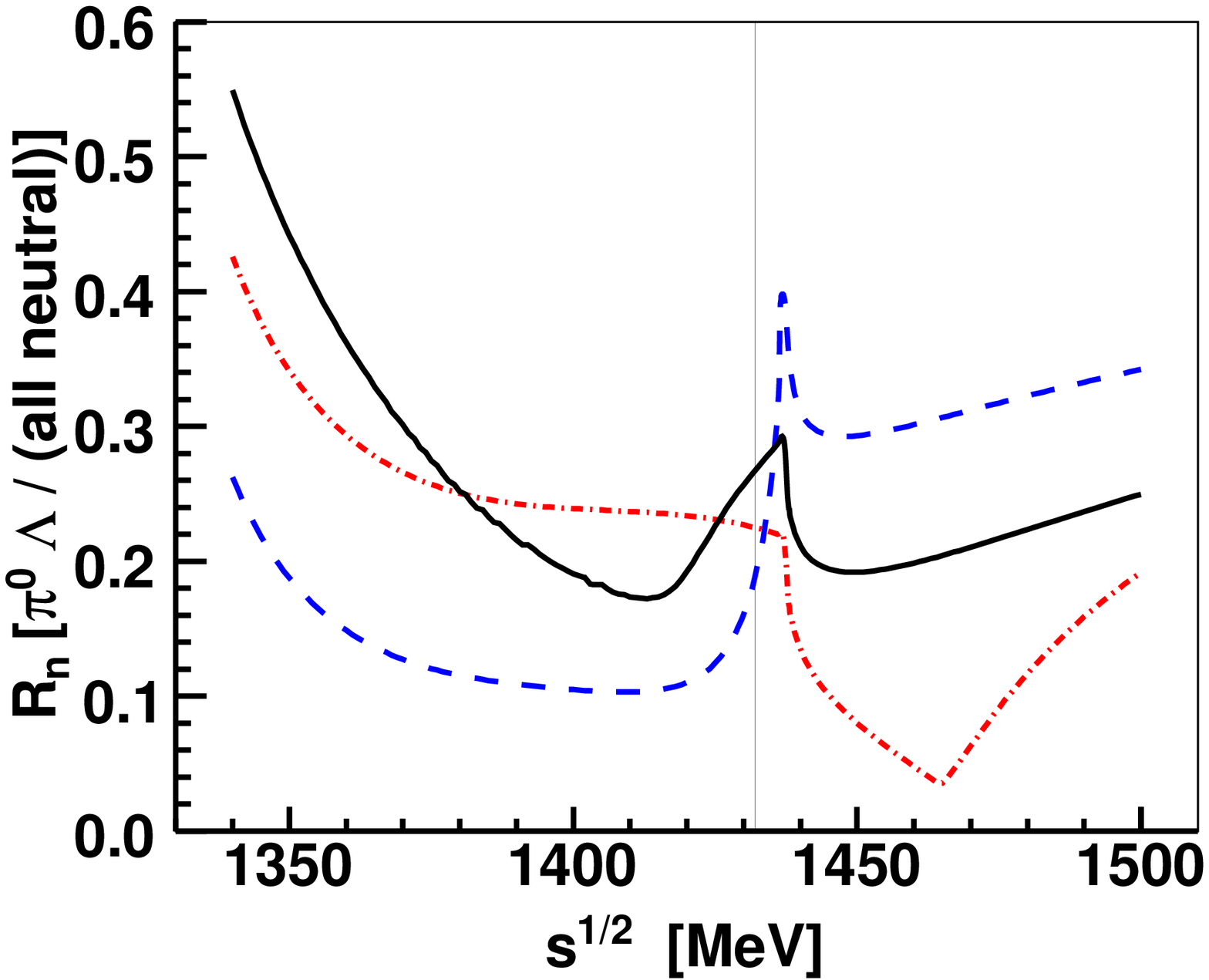}
\caption{Energy dependence of the branching ratios 
$R_c = \Gamma(\hbox{charged particles})/\Gamma(\hbox{all})$ (left panel) 
and $R_n = \Gamma(\pi^0\Lambda)/ \Gamma(\hbox{all neutral states})$ (right panel). 
The dashed (blue) line shows the free space results, the dot-dashed (red) and continuous lines demonstrate 
the effects of including Pauli blocking only and Pauli blocking together with hadron selfenergies, respectively.
The thin vertical line marks the $K^{-}p$ threshold.}
\label{fig:BR23}
\end{figure}

For a reference we also show the energy dependence of the other two branching ratios used in fits 
of the $K^{-}p$ experimental data. They are given in Fig.~\ref{fig:BR23} where the impact of nuclear medium 
on the ratios is demonstrated as well as their energy dependence. Although the opening of the $\bar{K}^{0}n$ 
channel has a significant impact on both branching ratios the scale of variations is much smaller than 
in the $\gamma_p$ case. This seems natural as the energy dependence of the probabilities present in the nominator 
and denominator is similar and to a large extent cancels out for both rates represented by the $R_c$ and $R_n$ 
branching ratios. It is interesting that the nuclear medium does not affect much the threshold values 
of both rates despite having an observable effect on their energy dependence, particularly in the $R_n$ case.

\begin{figure}[htb]
\includegraphics[scale=0.5]{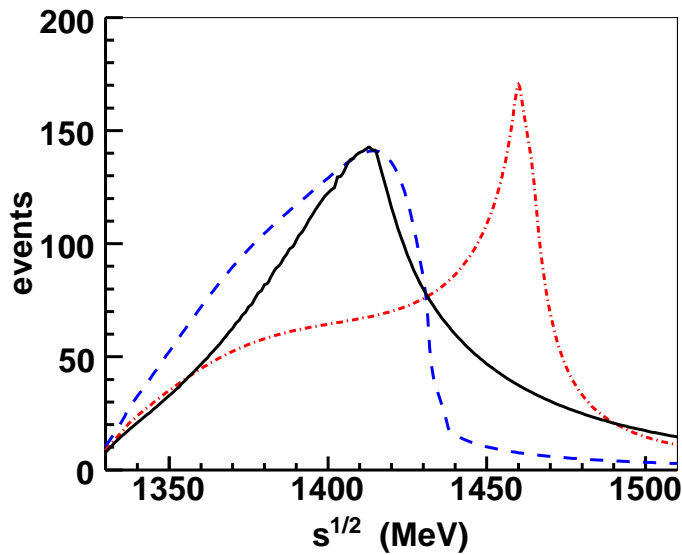}
\caption{$\pi\Sigma$ mass distribution for the reaction $K^{-}p \rightarrow \pi^{0}\Sigma^{0}$.  
 The dashed (blue) lines show the free space results, 
the dot-dashed (red) and continuous lines demonstrate the effects of including Pauli blocking only and Pauli blocking together with hadron selfenergies, respectively.}
\label{fig:pisig1}
\end{figure}

Finally, we look at the impact of nuclear medium on the shape of the $\pi\Sigma$ mass distributions. 
In Fig.~\ref{fig:pisig1} we present the mass distribution generated for the $K^{-}p \rightarrow \pi^{0}\Sigma^{0}$ 
reaction. In the free space the peak is located approximately at $1420$ MeV since the initial $K^{-}p$ channel 
couples more strongly to the higher (in terms of $\Re z$) of the two poles assigned to the $\Lambda(1405)$. 
The contribution of the lower pole and its distance from the real axis make the peak quite broad and 
clearly different from a Gaussian form. As we have already discussed the Pauli blocking moves the resonant 
structure to higher energies and the inclusion of hadron selfenergies moves it back to about its free-space position. 
Additionally, the in-medium peak is narrower and resembles more a typical Breit-Wigner distribution.
We also note that a larger imaginary part of the pion selfenergy would make the in-medium peak lower and slightly 
broader than the one computed for i$V^{\pi}_{0} = 30$ MeV and vice versa. Thus, the situation is similar 
to the one discussed in Fig.~\ref{fig:vpi} and the exact shape of the in-medium $\pi\Sigma$ mass distribution 
depends to some extend on the strength of pion absorption in nuclear matter.

\section{Conclusions}
\label{sec:conc}

We have presented a simplified model of meson-baryon interactions that is motivated by a chiral symmetry 
and large $N_c$ properties of QCD. It describes quite well the $S$-wave and $P$-wave experimental 
data from low energy $K^{-}p$ reactions. 
We have concentrated on implementing only the major driving forces of the meson-baryon interactions 
at energies around the $\pi\Sigma$ and $\bar{K}N$ thresholds, the leading order Tomozawa-Weinberg 
interaction and the direct formation of $\Sigma(1385)$ in the s-channel for the $S$-wave and 
$P$-wave interactions, respectively. We believe the simplicity of the model makes it suitable to study 
the effects caused by nuclear medium, particularly on the shape of the resonant $\pi\Sigma$ mass 
distributions that can be observed not only in the free space but in nuclear reactions as well. 
Understandably, the inclusion of NLO contributions in the meson-baryon interaction kernel as well as 
consideration of other $P$-wave mechanisms may alter our results to some extent. However, this would also 
add to a complexity of the theoretical model and most likely increase the ambiguity of its results. 
Still, there are opportunities for further development following this direction in the future.
We also note that processes not accounted for in the two-body meson-baryon framework represent 
additional theoretical limitations of the model. Particularly, one should anticipate slightly 
larger widths of the $\Lambda(1405)$ and $\Sigma(1385)$ resonances due to their decays 
into channels not included in the model or due to a larger than the adopted pion absorption 
in nuclear matter.

The energy and density dependence of $\gamma_p$ shown in Fig.~\ref{fig:BR1} can serve as a first step 
in understanding the observed results of the $\pi^{-}\Sigma^{+}$ to $\pi^{+}\Sigma^{-}$ formation rates 
reported for hypernuclear decays by the FINUDA collaboration \cite{2011FINUDA}. The $\gamma_p$ and $\gamma_n$ branching 
ratios are also relevant for an analysis of data on $K^{-}$ interactions with light nuclear targets 
that are measured by the AMADEUS collaboration in Frascati \cite{2013AMADEUS}. Evidently, a proper approach to these 
reactions is a more complex matter that requires a realistic treatment of the process dynamics as well as 
of the nuclear structure.

\vskip0.3in
{\bf \emph{Acknowledgments}}

We acknowledge a discussion with E.~Friedman and A.~Gal on the pion-nuclear optical potential.
The work of A.~C. was supported by the Grant Agency of the Czech Republic, Grant No.~15-04301S. 
V.~K. is supported by RIKEN programs for junior scientists.

\section*{Appendix --- $SU(3)$ Clebsch-Gordan coefficients and coupling matrices}

The ``${\rm meson}+{\rm baryon}\rightarrow{\rm baryon}$'' interaction vertex of Eq.~(\ref{eq:MBBvertex})
is determined (in the leading order) by two sets of Clebsch-Gordan coefficients: $SU(2)-{\rm spin}$
determining the spin-angular momentum transition, and $SU(3)$-flavor determining
baryon type transition. The $SU(3)$-flavor part is discussed here, the details of the $SU(2)-{\rm spin}$
can be found, for example, in Ref.~\cite{2012CK}.

From the group theory viewpoint, the interaction of octet meson with octet baryon
can be represented as a direct product of two octet representations. The key task is to find the decomposition
of this product into irreducible representations. The standard result is
\begin{equation}
\mathbf{8} \, \otimes \, \mathbf{8} = \mathbf{1} \, \oplus  
                                      \mathbf{8} \, \oplus
                                      \mathbf{8} \, \oplus
                                      \mathbf{10} \, \oplus
                                      \mathbf{\overline{10}} \, \oplus
                                      \mathbf{27} \;.
\label{eq:irrep}                                   
\end{equation}
One sees that there are two possibilities how to obtain an octet representation in the decomposition --- this fact
translates into the existence of two independent coefficients $D$ and $F$ in the conventional
$SU(3)$ chiral perturbation theory (where only octet baryons are considered).
On the other hand, there is just one decuplet representation in the decomposition of Eq. (\ref{eq:irrep}).
In principle, each transition carries an independent coupling constant
--- for now, let us label them $d_8$, $f_8$, and $g_{10}$ for the two octet and one decuplet representations, respectively.
The absolute values of $d_8$ and $f_8$ are determined by comparison to experiment (or to conventional $\chi$PT).
In order to determine $g_{10}$ we follow the large $N_c$ limit of QCD and its consistency relations stating
that pion-nucleon scattering amplitude must vanish in the leading order of the $1/N_c$ expansion.
Then, the corresponding values are:
\begin{align}
d_8 \,\, &\rightarrow \,\, \sqrt{\frac{5}{3}} D \;,\\
f_8 \,\, &\rightarrow \,\, \sqrt{3} F  \;,\\
\left(g_{10}\right)^2 \,\, &\rightarrow \,\, \frac{3}{4} \sqrt{\frac{2}{5}} (D+F)^2 = \frac{3}{4} \sqrt{\frac{2}{5}} \left(g_A\right)^2 \;.
\end{align} 

The individual Clebsch-Gordan coefficients relevant for our model are summarized in Table~\ref{tab:CGcoef}.
The inverse Clebsch-Gordan coefficient (corresponding to the decuplet-octet transition) is related to the
original one through a symmetry coefficient $\left(-2/\sqrt{5}\right)$.
The coupling matrix $\mathcal{C}^{d10}_{\Sigma^{*0}}$ determining the inter-channel coupling
is shown in Table~\ref{tab:Cmatrix}.

\begin{table}
\caption{Clebsch-Gordan coefficients for the octet-decuplet transition (baryon+meson to $\Sigma^{*0}$).}
\label{tab:CGcoef}
\begin{tabular}{|c|c|c|c|c|c|c|c|c|c|}
\hline
 $\pi ^0\Lambda$  & $\pi ^0\Sigma ^0$ & $\pi ^-\Sigma ^+$ & $\pi ^+\Sigma ^-$ & $K^-p$ &
 $\bar{K}^0n$ & $\eta \Lambda$  & $\eta \Sigma ^0$ & $K^0\Xi ^0$ & $K^+\Xi ^-$ 
\\ \hline
 $-\frac{1}{2}$ & $0$ & $-\frac{1}{2 \sqrt{3}}$ & $\frac{1}{2 \sqrt{3}}$ & $\frac{1}{2 \sqrt{3}}$ &
 $-\frac{1}{2 \sqrt{3}}$ & $0$ & $\frac{1}{2}$ & $\frac{1}{2 \sqrt{3}}$ & $-\frac{1}{2 \sqrt{3}}$ 
\\ \hline
\end{tabular}
\end{table}

\begin{widetext}
\begin{table}

\centering
\caption{Coupling matrix for the $P$-wave interaction $\mathcal{C}^{d10}_{\Sigma^{*0}}$.}
\label{tab:Cmatrix}
$
\begin{array}{|c|c|c|c|c|c|c|c|c|c|c|}
\hline
   & \pi ^0\Lambda  & \pi ^0\Sigma ^0 & \pi
   ^-\Sigma ^+ & \pi ^+\Sigma ^- & K^-p & \bar{K}^0n & \eta \Lambda  & \eta \Sigma ^0 &
   K^0\Xi ^0 & K^+\Xi ^- \\
\hline
 \pi ^0\Lambda  & -\frac{3 g_A^2}{8 \sqrt{2}} & 0 & -\frac{1}{8} \sqrt{\frac{3}{2}}
   g_A^2 & \frac{1}{8} \sqrt{\frac{3}{2}} g_A^2 & \frac{1}{8} \sqrt{\frac{3}{2}} g_A^2
   & -\frac{1}{8} \sqrt{\frac{3}{2}} g_A^2 & 0 & \frac{3 g_A^2}{8 \sqrt{2}} &
   \frac{1}{8} \sqrt{\frac{3}{2}} g_A^2 & -\frac{1}{8} \sqrt{\frac{3}{2}} g_A^2 \\
\hline
 \pi ^0\Sigma ^0 & \text{} & 0 & 0 & 0 & 0 & 0 & 0 & 0 & 0 & 0 \\
\hline
 \pi ^-\Sigma ^+ & \text{} & \text{} & -\frac{g_A^2}{8 \sqrt{2}} & \frac{g_A^2}{8
   \sqrt{2}} & \frac{g_A^2}{8 \sqrt{2}} & -\frac{g_A^2}{8 \sqrt{2}} & 0 & \frac{1}{8}
   \sqrt{\frac{3}{2}} g_A^2 & \frac{g_A^2}{8 \sqrt{2}} & -\frac{g_A^2}{8 \sqrt{2}} \\
\hline
 \pi ^+\Sigma ^- & \text{} & \text{} & \text{} & -\frac{g_A^2}{8 \sqrt{2}} &
   -\frac{g_A^2}{8 \sqrt{2}} & \frac{g_A^2}{8 \sqrt{2}} & 0 & -\frac{1}{8}
   \sqrt{\frac{3}{2}} g_A^2 & -\frac{g_A^2}{8 \sqrt{2}} & \frac{g_A^2}{8 \sqrt{2}} \\
\hline
 K^-p & \text{} & \text{} & \text{} & \text{} & -\frac{g_A^2}{8 \sqrt{2}} &
   \frac{g_A^2}{8 \sqrt{2}} & 0 & -\frac{1}{8} \sqrt{\frac{3}{2}} g_A^2 &
   -\frac{g_A^2}{8 \sqrt{2}} & \frac{g_A^2}{8 \sqrt{2}} \\
\hline
 \bar{K}^0n & \text{} & \text{} & \text{} & \text{} & \text{} & -\frac{g_A^2}{8
   \sqrt{2}} & 0 & \frac{1}{8} \sqrt{\frac{3}{2}} g_A^2 & \frac{g_A^2}{8 \sqrt{2}} &
   -\frac{g_A^2}{8 \sqrt{2}} \\
\hline
 \eta \Lambda  & \text{} & \text{} & \text{} & \text{} & \text{} & \text{} & 0 & 0 & 0
   & 0 \\
\hline
 \eta \Sigma ^0 & \text{} & \text{} & \text{} & \text{} & \text{} & \text{} & \text{} &
   -\frac{3 g_A^2}{8 \sqrt{2}} & -\frac{1}{8} \sqrt{\frac{3}{2}} g_A^2 & \frac{1}{8}
   \sqrt{\frac{3}{2}} g_A^2 \\
\hline
 K^0\Xi ^0 & \text{} & \text{} & \text{} & \text{} & \text{} & \text{} & \text{} &
   \text{} & -\frac{g_A^2}{8 \sqrt{2}} & \frac{g_A^2}{8 \sqrt{2}} \\
\hline
 K^+\Xi ^- & \text{} & \text{} & \text{} & \text{} & \text{} & \text{} & \text{} &
   \text{} & \text{} & -\frac{g_A^2}{8 \sqrt{2}} \\
\hline
\end{array}
$
\end{table}
\end{widetext}


\begin{thebibliography}{99}

\bibitem{1979Wei}
  S.~Weinberg,
  Physica 96A (1979) 327--340.
\bibitem{1985GL}
  J.~Gasser, H.~Leutwyler,
  Nucl.~Phys.~B 250 (1985) 465.
\bibitem{1995KSWnpa}
  N.~Kaiser, P.\,B.~Siegel, W.~Weise,
  Nucl.~Phys.~A 594 (1995) 325--345.
\bibitem{1998OR}
  E.~Oset and A.~Ramos, 
  Nucl.~Phys.~A 635 (1998) 99.
\bibitem{2013MAY}
  S.~Maeda, Y.~Akaishi, T.~Yamazaki, 
  Proc.~Jpn.~Acad., B 89 (2013) 418--437.
\bibitem{2006MFG}
  J.~Mare\v{s}, E.~Friedman, A.~Gal, 
  Nucl.~Phys.~A 770 (2006) 84.
\bibitem{2012GM}
  D.~Gazda, J.~Mare\v{s}, 
  Nucl.~Phys.~A 881 (2012) 159.
\bibitem{2008SSS}
  J.~Schaffner-Bielich, S.~Schramm, H.~St\"{o}cker, 
  Proc. Int’l.~School of Physics “Enrico Fermi”, Course CLXVII, Eds.~M.~Anselmino {\it et al.}
  (IOS Press, Amsterdam, 2008) 119–144.
\bibitem{1974tH} 
  G. t'Hooft, 
  Nucl.~Phys.~B 72 (1974) 461.
\bibitem{1979Wit} 
  E. Witten, 
  Nucl.~Phys.~B 160 (1979) 57.
\bibitem{1984GSprl} 
  J.L.~Gervais, B.~Sakita, 
  Phys.~Rev.~Lett.~52 (1984) 87.
\bibitem{1993DM-1}
  R.F.~Dashen, A.V.~Manohar, 
  Phys.~Lett.~B 513 (1993) 425.
\bibitem{2012HJ}
  T.~Hyodo, D.~Jido,
  Prog.~Part.~Nucl.~Phys.~67 (2012) 55--98.
\bibitem{2008HW}
  T.~Hyodo, W.~Weise,
  Phys.~Rev.~C 77 (2008) 035204.
\bibitem{2001OM}
  J.\,A.~Oller, U.-G.~Mei{\ss}ner, 
  Phys.~Lett.~B 500 (2001) 263--272. 
\bibitem{2003JOORM}
  D.~Jido, J.\,A.~Oller, E.~Oset, A.~Ramos, U.-G.~Mei{\ss}ner,
  Nucl.~Phys.~A~725 (2003) 181--200.
\bibitem{2013HADES}
  G.~Agakishievet {\it et al.} [HADES Collaboration],
  Phys.~Rev.~C 87 (2013) 025201.
\bibitem{2008ANKE}
  I.~Zychor {\it et al.} [ANKE Collaboration],
  Phys.~Lett.~B 660 (2008) 167--171.
\bibitem{1984Hem}
  R.\,J.~Hemingway,
  Nucl.~Phys.~B 253 (1985) 742--752.
\bibitem{2013CLAS}
  K.~Moriya {\it et al.} [CLAS Collaboration],
  Phys.~Rev.~C 88 (2013) 045201.
\bibitem{2013RO}
  L.~Roca, E.~Oset,
  Phys.~Rev.~C87, 055201 (2013).
\bibitem{2015MM}
  M.~Mai, U.-G.~Mei{\ss}ner,
  Eur.~Phys.~J.~A51, 30 (2015).
\bibitem{2013AMAD}
  A.~Scordo {\it et al.} [AMADEUS Collaboration],
  arXiv:1304.7149 [nucl-ex].
\bibitem{2005BNW}
  B.~Borasoy, R.~Ni{\ss}ler, W.~Weise,
  Eur.~Phys.~J.~A 25 (2005) 79--96.
\bibitem{2012CS}
  A.~Ciepl\'{y}, J.~Smejkal,
  Nucl.~Phys.~A 881 (2012) 115--126.
\bibitem{2012IHW}
  Y.~Ikeda, T.~Hyodo, W.~Weise, 
  Nucl.~Phys.~A 881 (2012) 98-114.
\bibitem{2013GO}
  Z.-H.~Guo, J.~Oller, 
  Phys.~Rev.~C 87 (2013) 035202.
\bibitem{1996WKW} 
  T.~Waas, N.~Kaiser, W.~Weise, 
  Phys.~Lett.~B 365 (1996) 12--16. 
\bibitem{1998Lut}
  M.~Lutz,
  Phys.~Lett.~B 426 (1998) 12--20.
\bibitem{2000RO}
  A.~Ramos, E.~Oset,
  Nucl.~Phys.~A 671 (2000) 481--502.
\bibitem{2001CFGM}
  A.~Ciepl\'{y}, E.~Friedman, A.~Gal, J.~Mare\v{s},
  Nucl.~Phys.~A 696 (2001) 173--193.
\bibitem{2006TRO}
  L.~Tol\'{o}s, A.~Ramos, E.~Oset,
  Phys.~Rev.~C 74, 015203 (2006).
\bibitem{2011CFGGM}
  A.~Ciepl\'{y}, E.~Friedman, A.~Gal, D.~Gazda, J.~Mare\v{s},
  Phys.~Rev.~C 84 (2011) 045206.
\bibitem{1994FGB}
  E.~Friedman, A.~Gal, C.\,J.~Batty,
  Nucl.~Phys.~A 579 (1994) 518--538.
\bibitem{2013FG}
  E.~Friedman, A.~Gal,
  Nucl.~Phys.~A 899 (2013) 60.
\bibitem{2002JOR}
  D.~Jido, E.~Oset, A.~Ramos,
  Phys.~Rev.~C 66 (2002) 055203.
\bibitem{2012Kre}
  V.~Krej\v{c}i\v{r}\'ik, 
  Phys.~Rev.~C 86 (2012) 024003. 
\bibitem{1964GW} 
  M.L.~Goldberger, K.M. Watson, 
  {\it Collision theory}, John Willey \& Sons (1964).
\bibitem{1954Yam} 
  Y.~Yamaguchi, 
  Phys.~Rev.~95 (1954) 1628.
\bibitem{1954YamYam} 
  Y.~Yamaguchi, Y.~Yamaguchi, 
  Phys.~Rev.~95 (1954) 1635.
\bibitem{1984GSprd} 
  J.L.~Gervais, B.~Sakita, 
  Phys.~Rev.~D 30 (1984) 1795.
\bibitem{1993DM-2} 
  R.F.~Dashen, A.V.~Manohar, 
  Phys.~Lett. B 513 (1993) 438.
\bibitem{1994DJM} 
  R.F.~Dashen, E.E.~Jenkins, A.V.~Manohar, 
  Phys.~Rev.~D 49 (1994) 4713.
\bibitem{1995DJM} 
  R.F.~Dashen, E.E. Jenkins, A.V. Manohar, 
  Phys.~Rev.~D 51 (1995) 3697.
\bibitem{2012CK} 
  T.D.~Cohen, V.~Krej\v{c}i\v{r}\'ik, 
  Phys.~Rev.~C 85 (2012) 035205. 
\bibitem{1963dS} 
  J.J.~de Swart, 
  Rev.~Mod.~Phys.~35 (1963) 16.
\bibitem{2004CL} 
  T.D.~Cohen, R.F.~Lebed, 
  Phys.~Rev.~D 70 (2004) 096015.
\bibitem{2014CFGM}
  A.~Ciepl\'{y}, E.~Friedman, A.~Gal, J.~Mare\v{s},
  Nucl.~Phys.~{\bf A925} 126 (2014).
\bibitem{1996JS}
  M.~B.~Johnson, G.~R.~Satchler, 
  Ann.~Phys.~248, 134 (1996).
\bibitem{SAID}
  Center of Nuclear Study Data Analysis Center, 
  URL: http://gwdac.phys.gwu.edu/KWW.
\bibitem{1983F} 
  E.~Friedman,
  Phys.~Rev.~C 28, 1264 (1983).
\bibitem{1982Cib} 
  J.~Ciborowski {\it et al.}, 
  J.~Phys. G 8 (1982) 13.
\bibitem{1983Ev} 
  D. Evans {\it et al.}, 
  J.~Phys. G 9 (1983) 885.
\bibitem{1965Sak} 
  M.~Sakitt {\it et al.}, 
  Phys.~Rev.~139 (1965) 719.
\bibitem{1962HR} 
  W.E.~Humphrey, R.R. Ross, 
  Phys.~Rev.~127 (1962) 1305.
\bibitem{1976Ma} 
  T.S.~Mast {\it et al.}, 
  Phys.~Rev. D 14 (1976) 13.
\bibitem{1981Ban} 
  R.O.~Bangerter {\it et al.}, 
  Phys.~Rev. D 23 (1981) 1485.
\bibitem{1963Wa} 
  M.B.~Watson, M.~Ferro-Luzzi, R.D. Tripp, 
  Phys.~Rev.~131 (1963) 2248.
\bibitem{2014PDG}
  K.A.~Olive {\it et al.} (Particle Data Group), 
  Chin.~Phys.~C 38, 090001 (2014). 
\bibitem{1981Mar}
  A.D.~Martin, 
  Nucl.~Phys.~B 179 (1981) 33.
\bibitem{2011FINUDA}
  M.~Agnelo {\it et al.} [FINUDA Collaboration], 
  Phys.~Lett.~B 704 (2011) 474-480.
\bibitem{2013AMADEUS}
  K.~Piscicchia {\it et al.} [AMADEUS Collaboration], 
  PoS Bormio2013 (2013) 034, arXiv:13044.7165 [nucl-ex].



\end{thebibliography}
\end{document}